\documentclass[12pt,twoside,pointlessnumbers,nofootinbib,smallheadings]{revtex4}

\usepackage[CJKbookmarks]{hyperref}
\usepackage{CJK}
\usepackage{amsmath}
\usepackage{graphics}
\usepackage{slashed}
\usepackage{graphicx}
\usepackage{indentfirst}
\usepackage{amssymb}
\usepackage{cancel}
\usepackage{leftidx}
\usepackage{simplewick}
\usepackage[page,title,titletoc,header]{appendix}

\begin{document}

\begin{CJK*}{GBK}{song}

\title{Higgs-$\mu$-$\tau$ Coupling at High and Low Energy Colliders}

\author{Ying-nan Mao $^1$ and Shou-hua
Zhu $^{1,2,3}$  }

\affiliation{
$ ^1$ Institute of Theoretical Physics $\&$ State Key Laboratory of
Nuclear Physics and Technology, Peking University, Beijing 100871,
China \\
$ ^2$ Collaborative Innovation Center of Quantum Matter, Beijing 100871, China \\
$ ^3$ Center for High Energy Physics, Peking University,
Beijing 100871, China
}

\begin{abstract}

There is no tree-level flavor changing neutral current (FCNC) in the standard model (SM) which contains
only one Higgs doublet. If more Higgs doublets are introduced for various reasons, the tree level FCNC would
be inevitable except extra symmetry was imposed. Therefore FCNC processes are the excellent probes for
the physics beyond the SM (BSM). In this paper,
we studied the lepton flavor violated (LFV) decay processes $h\rightarrow\mu\tau$ and $\tau\rightarrow\mu\gamma$
induced by Higgs-$\mu$-$\tau$ vertex. For $\tau\rightarrow\mu\gamma$, its branching ratio is also related to the
$ht\bar{t}$, $h\tau^+\tau^-$ and $hW^+W^-$ vertices. We categorized the BSM into two scenarios for the Higgs coupling strengths near
or away from SM. For the latter scenario, we took the spontaneously broken two Higgs doublet model (Lee model) as an example.
We considered the constraints by recent data from LHC and B factories, and found that the measurements gave weak constraints.
At LHC Run II,  $h\rightarrow\mu\tau$ will be confirmed or set stricter limit on its branching ratio.
Accordingly, $\textrm{Br}(\tau\rightarrow\mu\gamma)\lesssim\mathcal{O}(10^{-10}-10^{-8})$
for general chosen parameters. For the positive case, $\tau\rightarrow\mu\gamma$ can be discovered
with $\mathcal{O}(10^{10})$ $\tau$ pair samples at SuperB factory, Super $\tau$-charm factory and new Z-factory.
The future measurements for $\textrm{Br}(h\rightarrow\mu\tau)$ and $\textrm{Br}(\tau\rightarrow\mu\gamma)$ will be used
to distinguish these two scenarios or set strict constraints on the correlations among different Higgs couplings, please
see \autoref{summary} in the text for details.

\end{abstract}

\date{\today}

\maketitle

\newpage

\section{Introduction}
In the standard model (SM), we can diagonalize the gauge couplings and Yukawa couplings simultaneously,
i.e., there is no flavor changing neutral current (FCNC) at the tree level.
In the quark sector, flavor
changing neutral currents occur at loop level with the help of CKM quark mixing matrix \cite{CKM}.
However, in the lepton sector, it is extremely suppressed by GIM mechanism \cite{GIM} in the SM
due to the smallness of neutrino mass. For example, for the lepton flavor violation (LFV) process
$\ell_i\rightarrow\ell_j\gamma$
\begin{equation}
\frac{\Gamma(\ell_i\rightarrow\ell_j\gamma)}{\Gamma(\ell_i\rightarrow\ell_j\nu_i\bar{\nu}_j)}=
\frac{3\alpha}{32\pi}\left|\mathop{\sum}_kV_{ik}^*V_{jk}\frac{m^2_{\nu_k}}{m^2_W}\right|^2
\end{equation}
in the SM \cite{fcsm} where $V_{ij}$ are the PMNS lepton mixing matrix \cite{PMNS} elements. With the
data from neutrino oscillation \cite{no}, it is estimated to be
\begin{equation}
\textrm{Br}(\mu\rightarrow e\gamma)\sim\mathcal{O}(10^{-56})\quad\quad\textrm{and}
\quad\quad\textrm{Br}(\tau\rightarrow e(\mu)\gamma)\sim\mathcal{O}(10^{-55}-10^{-54})
\end{equation}
in the SM. It is far away from the recent experimental upper limit \cite{exp1,exp2}\footnote{For either B factory with
$L\approx0.5\textrm{ab}^{-1}$ luminosity at $\sqrt{s}=10.6\textrm{GeV}$ ($\Upsilon(4S)$ threshold).}
\begin{eqnarray}
&\textrm{Br}(\tau\rightarrow e\gamma)<\left\{\begin{array}{cc}1.2\times10^{-7}&(\textrm{Belle})\\3.3\times10^{-8}&(\textrm{BaBar})\end{array}\right.,
\quad\textrm{Br}(\tau\rightarrow\mu\gamma)<\left\{\begin{array}{cc}4.5\times10^{-8}&(\textrm{Belle})\\4.4\times10^{-8}&(\textrm{BaBar})\end{array}\right.,
\nonumber\\
&\textrm{and}\quad\textrm{Br}(\mu\rightarrow e\gamma)<5.7\times10^{-13}\quad(\textrm{MEG}),\quad\textrm{all at $90\%$ C.L.}
\end{eqnarray}
and the near future sensitivities with the improvement of an order \cite{fut1,fut2,fut3,fut4}. So that the discovery of the signals
$\ell_i\rightarrow\ell_j\gamma$ at future colliders would clearly indicate new physics (NP) beyond the SM (BSM).
Generally speaking the FCNC process will be one of the best probe of the BSM for future hadron and electron-positron colliders \cite{Zhu2014hda}.

In July 2012, a new boson was discovered at LHC \cite{dis1,dis2}, and its properties are like those of a SM Higgs boson \cite{pro}.
The Higgs mediated LFV process is attractive because of a $2.4\sigma$ hint found by CMS Collaboration \cite{CMS}
in the search for $h\rightarrow\mu\tau$ process\footnote{Recently the ATLAS Collaboration also published the searching
result in the same process \cite{ATLAS} with the result close to that in \cite{CMS} by CMS Collaboration.}.
Assuming that the Higgs production cross section and total decay width
are the same as those in the SM, the best fit (B.F.) branching ratio and $95\%$ upper limit (U.L.) are respectively
\cite{CMS}\footnote{For the full LHC Run I data with $L\approx25\textrm{fb}^{-1}$ luminosity at $\sqrt{s}=(7-8)\textrm{TeV}$.}
\begin{equation}
\label{res}
\textrm{Br}(h\rightarrow\mu\tau)=(0.84^{+0.39}_{-0.37})\%\quad(\textrm{B.F.})\quad\textrm{and}\quad\textrm{Br}(h\rightarrow\mu\tau)<1.51\% \quad(\textrm{U.L.})
\end{equation}
If this signature was confirmed at future colliders, it would clearly indicate NP in the Higgs sector.
In the extensions of SM, there may be direct Higgs-$\mu$-$\tau$ coupling to explain this hint, for example,
in some types of two Higgs doublet models (2HDM) \cite{2hdm}, like type III 2HDM \cite{2hdm3,2hdm5,2hdmanother,2hdmanother2}, 2HDMs
with other flavor symmetries \cite{2hdms,2hdms2,2hdms3}, Lee model \cite{our,Lee}, and other models \cite{other1,other2,other3}. It may be also
related to other phenomena like the excess in $t\bar{t}h$ searches \cite{phe1}, $b\rightarrow s$ semi-leptonic
decays \cite{2hdms2}, anomalous magnetic moment $(g-2)$ for $\mu$ \cite{phe2}, LFV $\tau$
decays \cite{2hdms3,phe2,2hdmanother,tau1,tau2}, or even the lepton flavored dark matter \cite{phe3}. Writing the Higgs-$\mu$-$\tau$ vertex as
\begin{equation}
\label{hmt}
\mathcal{L}_{h\mu\tau}=-\frac{h}{\sqrt{2}}\left(Y_{\mu\tau}\bar{\mu}_L\tau_R+Y_{\tau\mu}\bar{\tau}_L\mu_R+\textrm{h.c.}\right),
\end{equation}
and adopting the Cheng-Sher ansatz \cite{CS}, the data gave \cite{CMS}
\begin{equation}
\label{resc}
\sqrt{|Y_{\mu\tau}|^2+|Y_{\tau\mu}|^2}<5\times10^{-3}\quad\textrm{or}\quad
\sqrt{\frac{(|Y_{\mu\tau}|^2+|Y_{\tau\mu}|^2)v^2}{2m_{\mu}m_{\tau}}}<2.
\end{equation}

In the future, at low energy $e^+e^-$ colliders like Super-B factory \cite{fut2,fut3}, Super $\tau$-charm factory \cite{STC,STC2} or
the new Z-factory \cite{zfactory},
there would be signatures or stricter constraints for $\tau\rightarrow\mu\gamma$ process;
and at high energy colliders like LHC Run II at
$\sqrt{s}=(13-14)\textrm{TeV}$, there would be signatures or stricter constraints for $h\rightarrow\mu\tau$ process. The
results would be comparable and may give new constraints on the Higgs-$\mu$-$\tau$ coupling or the correlations among
the couplings between Higgs and other particles.

This paper is organized as follows. In \autoref{IADW} we present the
effective interactions and branching ratios for $h\rightarrow\mu\tau$ and $\tau\rightarrow\mu\gamma$ processes; \autoref{CBRE}
and \autoref{CAFC} contain the constraints from recent data and at future colliders respectively; \autoref{CAD} are
our conclusions and discussions.

\section{Effective Higgs-$\mu$-$\tau$ Interaction and Decay Widths for $h\rightarrow\mu\tau$ and $\tau\rightarrow\mu\gamma$ Processes}
\label{IADW}
Based on 2HDM (type III), the higgs effective couplings can be written as
\begin{eqnarray}
\label{lar}
\mathcal{L}_{h}&=&c_Vh\left(\frac{2m^2_W}{v}W^{+\mu}W^-_{\mu}+\frac{m^2_Z}{v}Z^{\mu}Z_{\mu}\right)
-h\left(\frac{c_tm_t}{v}\bar{t}_Lt_R+\frac{c_{\tau}m_{\tau}}{v}\bar{\tau}_L\tau_R+\textrm{h.c.}\right)\nonumber\\
&&-\frac{h}{\sqrt{2}}\left(Y_{\mu\tau}\bar{\mu}_L\tau_R+Y_{\tau\mu}\bar{\tau}_L\mu_R+\textrm{h.c.}\right)
\end{eqnarray}
where $c_i$ stands for the coupling strength ratio compared with that in SM\footnote{The $c_t$ and $c_{\tau}$ may be
complex while $c_V$ must be real, and in the SM $c_V=c_t=c_{\tau}=1$.} and $Y_{ij}$ stands for the LFV coupling
just like that in Eq. (\ref{hmt}). With a direct calculation \cite{CMS}, for the $h\rightarrow\mu\tau$ process, we have
\begin{equation}
\textrm{Br}(h\rightarrow\mu\tau)=\frac{m_h}{16\pi\Gamma_h}\left(|Y_{\mu\tau}|^2+|Y_{\tau\mu}|^2\right)
\end{equation}
where $\Gamma_h$ means the total decay width of Higgs boson and in SM we have $\Gamma_{h,\textrm{SM}}=4.1\textrm{MeV}$
\cite{width} for $m_h=125\textrm{GeV}$.

\begin{figure}
\caption{Typical Feynman diagrams which contribute to $\tau\rightarrow\mu\gamma$ decay. The left one is Higgs-mediated
one-loop diagram, the middle and right ones are $W$ boson and top quark mediated Barr-Zee Type two-loop diagrams
respectively.}\label{Fey}
\includegraphics[scale=0.7]{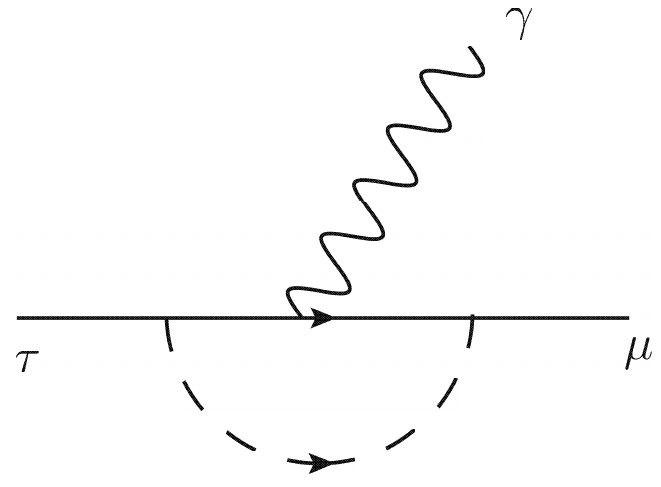}\quad\includegraphics[scale=0.7]{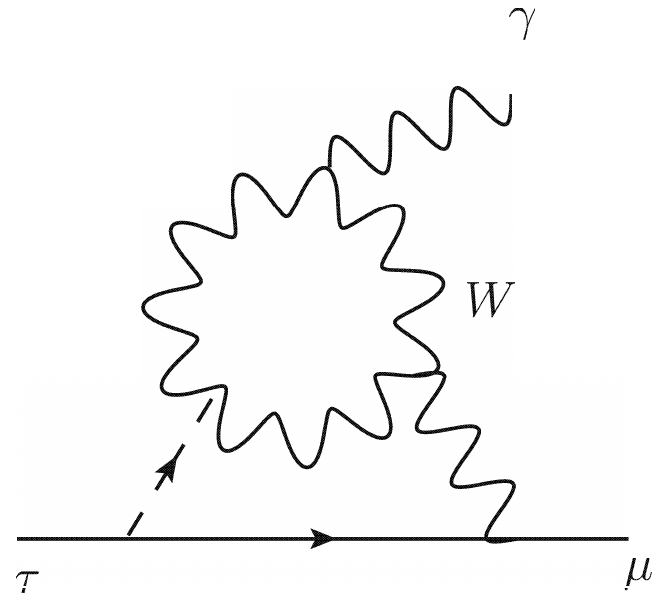}\quad\includegraphics[scale=0.7]{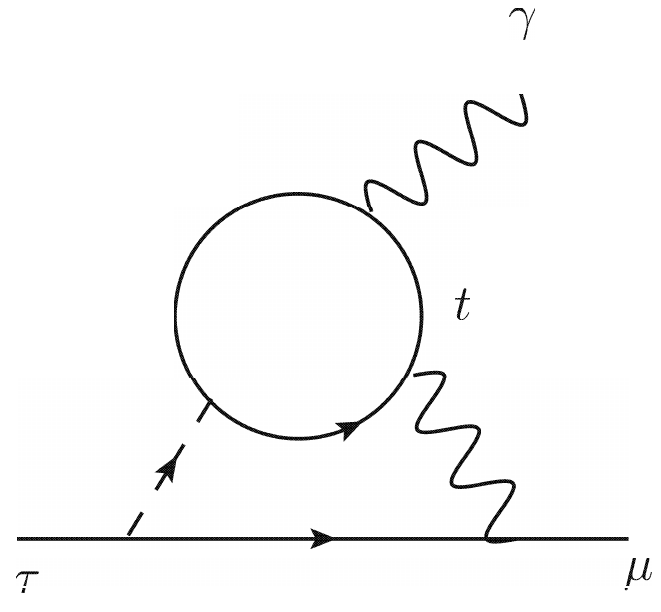}
\end{figure}

The $\tau\rightarrow\mu\gamma$ decay process is loop induced. The dominant contribution usually comes from Barr-Zee
type\footnote{This type of two-loop diagrams was first proposed by Barr and Zee \cite{BZ} during the calculation for
lepton electric dipole moment (EDM).} two-loop diagrams since there is an additional $(m_{\tau}/m_h)^2\log(m^2_h/m^2_{\tau})\sim\mathcal{O}(10^{-3})$
suppression in the one-loop amplitude \cite{loop}, see the Feynman diagrams in \autoref{Fey}.
Following the formulae in \cite{loop},
\begin{equation}
\label{ratio}
\frac{\textrm{Br}(\tau\rightarrow\mu\gamma)}{\textrm{Br}(\tau\rightarrow\mu\nu\bar{\nu})}=\frac{48\pi^3\alpha}{G^2_F}
\left(|\mathcal{A}_L|^2+|\mathcal{A}_R|^2\right).
\end{equation}
Here the left (right) handed amplitudes $\mathcal{A}_{L(R)}$ can be expressed as \cite{2hdm5,loop,loop2,phe2}
\footnote{The results in these papers are different. We checked the calculations and got the result
consistent with that in \cite{phe2} by Omura et. al.}
\begin{eqnarray}
&&\mathcal{A}_{L}(\mathcal{A}^*_R)\nonumber\\
&=&\mathcal{A}_{L,\textrm{1-loop}}(\mathcal{A}^*_{R,\textrm{1-loop}})
+\mathcal{A}_{L,\textrm{2-loop}}(\mathcal{A}^*_{R,\textrm{2-loop}})\nonumber\\
&=&\frac{Y_{\mu\tau}(Y^*_{\tau\mu})}{16\sqrt{2}\pi^2}\Bigg(\frac{m_{\tau}}{m^2_hv}
\left(c_{\tau}\ln\left(\frac{m^2_h}{m^2_{\tau}}\right)-\frac{4}{3}\textrm{Re}(c_{\tau})-\frac{5i}{3}\textrm{Im}(c_{\tau})\right)\nonumber\\
&&+\frac{c_V\alpha}{\pi m_{\tau}v}\left(\left(3+\frac{m^2_h}{2m^2_W}\right)f\left(\frac{m^2_W}{m^2_h}\right)+
\left(\frac{23}{4}-\frac{m^2_h}{2m^2_W}\right)g\left(\frac{m^2_W}{m^2_h}\right)+\frac{3}{4}h\left(\frac{m^2_W}{m^2_h}\right)\right)\nonumber\\
\label{amp}
&&-\left.\frac{8\alpha}{3\pi m_{\tau}v}
\left(\textrm{Re}(c_t)f\left(\frac{m^2_t}{m^2_h}\right)+i\textrm{Im}(c_t)g\left(\frac{m^2_t}{m^2_h}\right)\right)\right)
\end{eqnarray}
where all the functions $f,g$ and $h$ come from 2-loop integrations as \cite{loop2}
\begin{eqnarray}
f(z)&=&\frac{z}{2}\int_0^1dx\frac{1-2x(1-x)}{x(1-x)-z}\ln\left(\frac{x(1-x)}{z}\right);\\
g(z)&=&\frac{z}{2}\int_0^1dx\frac{1}{x(1-x)-z}\ln\left(\frac{x(1-x)}{z}\right);\\
h(z)&=&-\frac{z}{2}\int_0^1dx\frac{1}{x(1-x)-z}\left(1-\frac{z}{x(1-x)-z}\ln\left(\frac{x(1-x)}{z}\right)\right).
\end{eqnarray}
The small contributions from heavy neutral higgses, charged higgs and $Z$-mediated loop are all ignored. Defining
\begin{equation}
\mathcal{A}\equiv\frac{\mathcal{A}_L}{Y_{\mu\tau}}=\frac{\mathcal{A}_R}{Y_{\tau\mu}},
\end{equation}
the equation (\ref{ratio}) should be changed to
\begin{equation}
\frac{\textrm{Br}(\tau\rightarrow\mu\gamma)}{\textrm{Br}(\tau\rightarrow\mu\nu\bar{\nu})}=\frac{48\pi^3\alpha|\mathcal{A}|^2}{G^2_F}
\left(|Y_{\mu\tau}|^2+|Y_{\tau\mu}|^2\right).
\end{equation}
Here $\textrm{Br}(\tau\rightarrow\mu\nu\bar{\nu})=17.4\%$ from PDG \cite{no}.

For both decay processes, the LFV parameter comes in the form $\sqrt{|Y_{\mu\tau}|^2+|Y_{\tau\mu}|^2}$, thus we do not
need to study the details about the chiral properties of the LFV coupling. Since both
$\textrm{Br}\propto(|Y_{\mu\tau}|^2+|Y_{\tau\mu}|^2)$, the ratio
$\textrm{Br}(\tau\rightarrow\mu\gamma)/\textrm{Br}(h\rightarrow\mu\tau)$ does not depend on $Y_{\mu\tau(\tau\mu)}$.
Therefore in this paper we will focus on the correlations among the Higgs couplings.

\section{Constraints by Recent Experiments}
\label{CBRE}
In general cases, $\alpha_t\equiv\arg(c_t)$ and $\alpha_{\tau}\equiv\arg(c_{\tau})$ may be nonzero. The replacement
\begin{equation}
\label{rep}
\textrm{Br}(h\rightarrow\mu\tau)\rightarrow\frac{\sigma_h}{\sigma_{h,\textrm{SM}}}\textrm{Br}(h\rightarrow\mu\tau)
=(\cos^2\alpha_t+2.31\sin^2\alpha_t)\textrm{Br}(h\rightarrow\mu\tau)
\end{equation}
should also be taken into account in (\ref{res}) where $\sigma_h$ stands for the Higgs production cross section\footnote{Gluon fusion
process is dominant in this case.} and $\sigma_{h,\textrm{SM}}$ means that in SM. To consider the numerical constraints on the couplings
in (\ref{lar}), we should take some benchmark points. Our fitting results \cite{our} preferred $|c_{\tau}|\sim1$ for almost all chosen
for other parameters, so in this paper we take $|c_{\tau}|=1$. The regions $c_V\lesssim0.4$, $|c_t|\lesssim0.5$ and $|c_t|\gtrsim2$
are excluded for most cases by our fitting results, so we never consider those regions in this paper.

\begin{figure}
\caption{Distribution for $R\equiv\textrm{Br}(\tau\rightarrow\mu\gamma)/\textrm{Br}(h\rightarrow\mu\tau)$ in $c_V-|c_t|$ plane
in unit of $(\Gamma_{h,\textrm{tot}})/\textrm{MeV})$ fixing $c_{\tau}=1$. We take $\alpha_t=(0,\pi/10,\pi/6)$ from left to right.
The green regions are for $R<10^{-10}$; the yellow regions are for $10^{-10}\leq R<10^{-9}$; the blue regions are for
$10^{-9}\leq R<10^{-8}$; the cyan regions are for $10^{-8}\leq R<3\times10^{-8}$; the orange regions are for $3\times10^{-8}\leq
R<6\times10^{-8}$; the red regions are for $6\times10^{-8}\leq R<10^{-7}$; and the brown regions are for
$10^{-7}\leq R<1.5\times10^{-7}$.}\label{cvct1}
\includegraphics[scale=0.4]{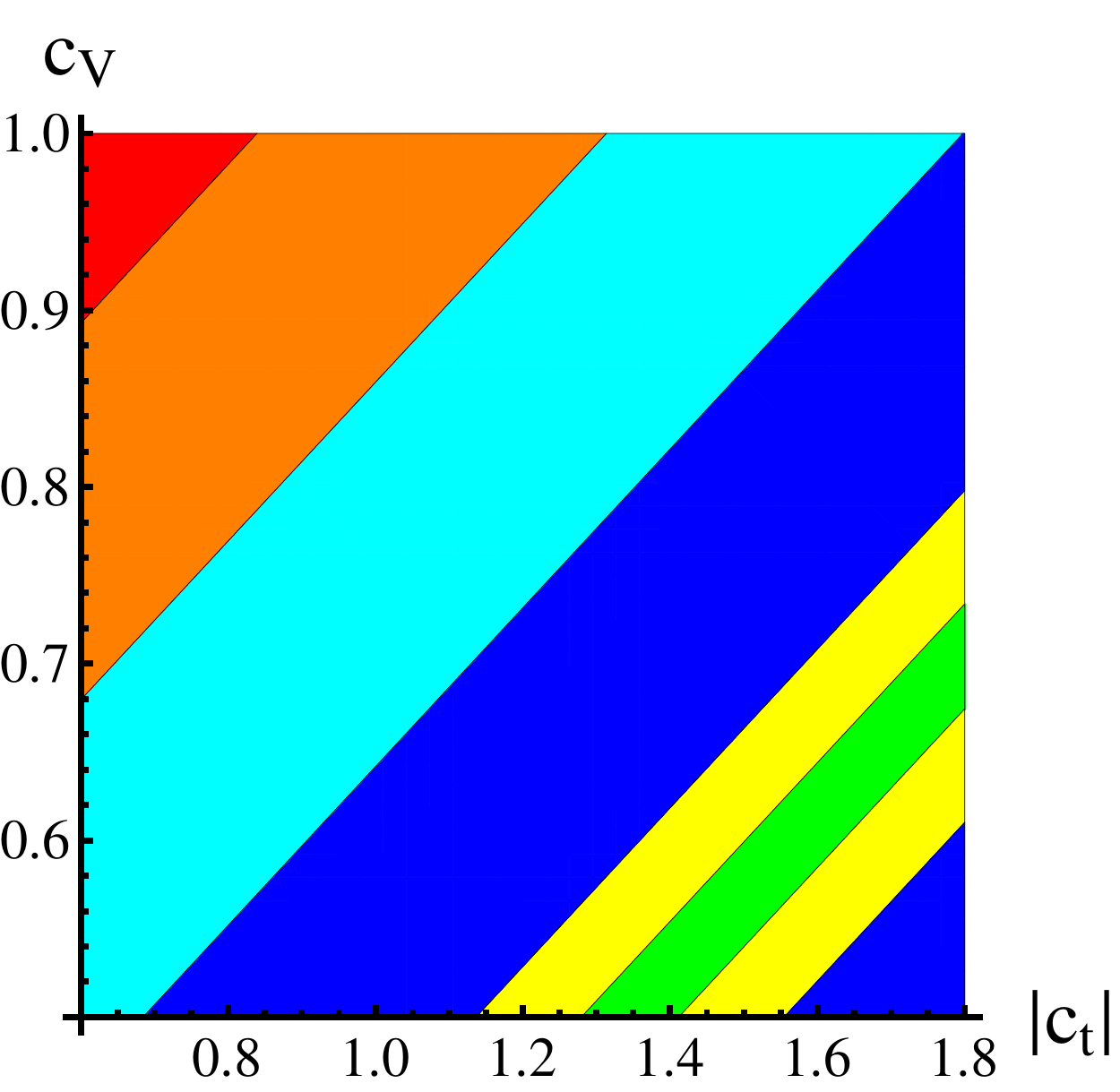}\quad\includegraphics[scale=0.4]{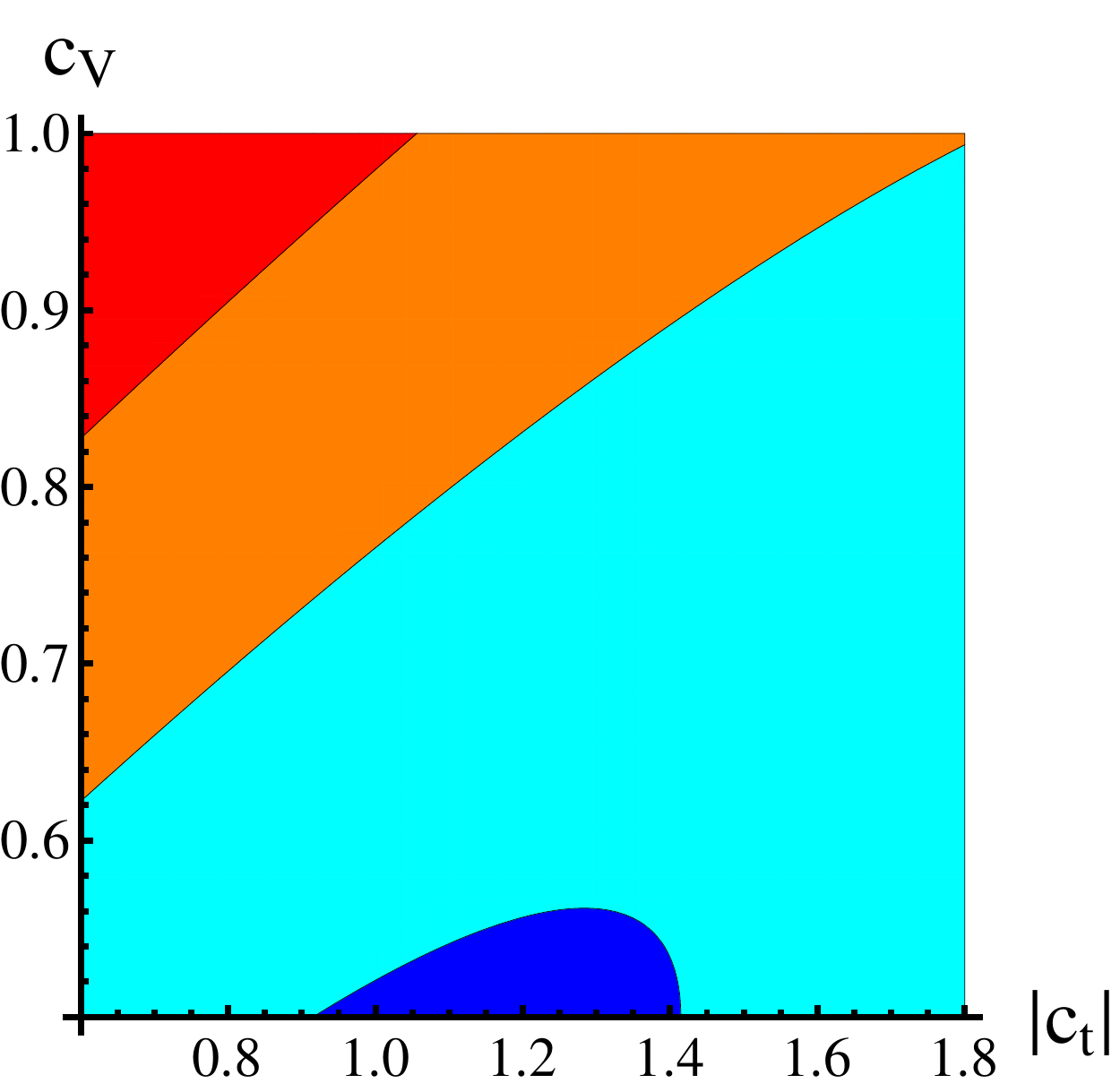}\quad\includegraphics[scale=0.4]{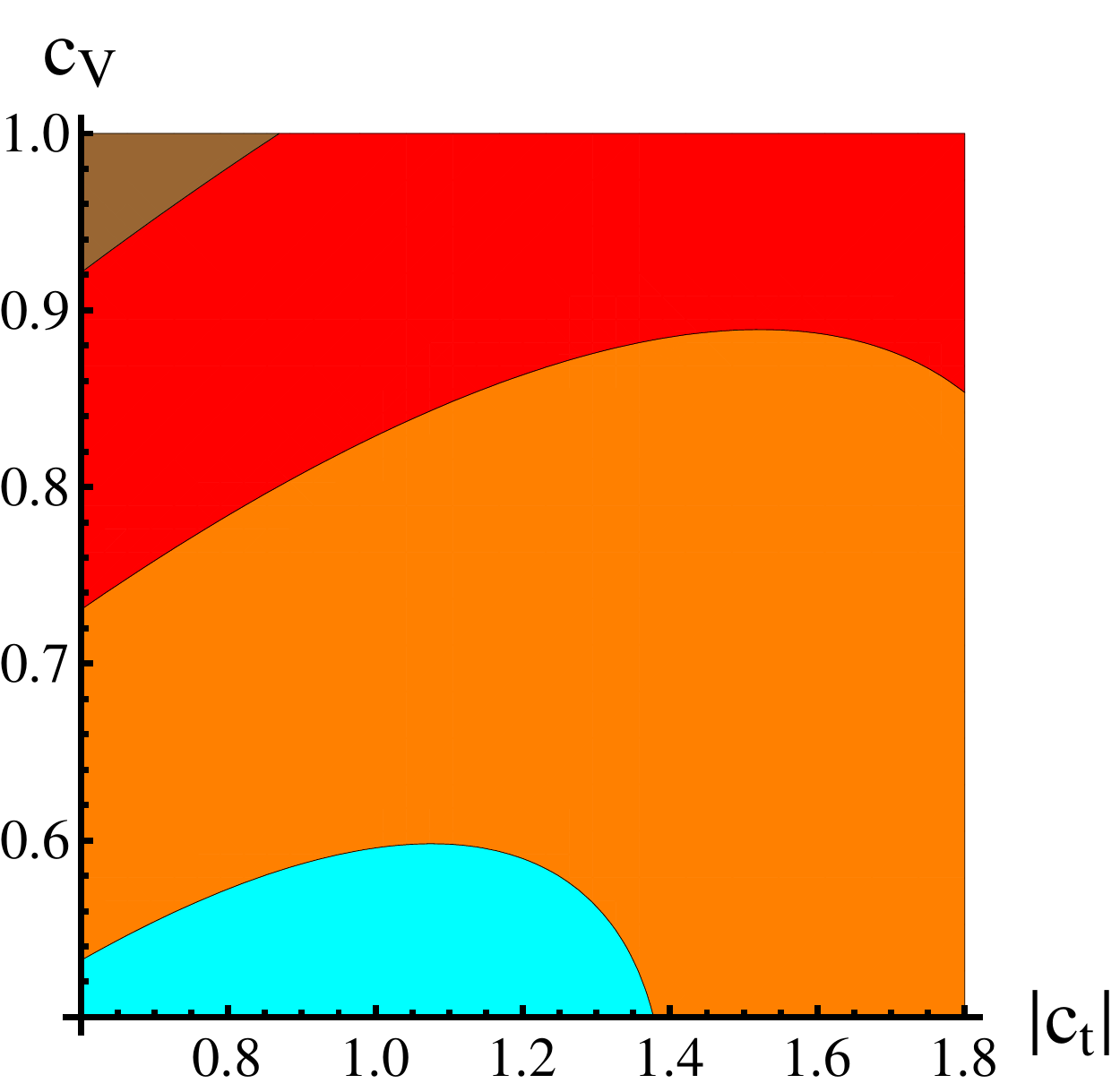}
\end{figure}

\begin{figure}
\caption{Distribution for $R\equiv\textrm{Br}(\tau\rightarrow\mu\gamma)/\textrm{Br}(h\rightarrow\mu\tau)$ in $c_V-|c_t|$ plane
in unit of $(\Gamma_{h,\textrm{tot}})/\textrm{MeV})$ fixing $c_{\tau}=1$. We take $\alpha_t=(\pi/4,\pi/2,2\pi/3)$ from left to right.
The green regions are for $R<10^{-7}$; the yellow regions are for $10^{-7}\leq R<2\times10^{-7}$; the blue regions are for
$2\times 10^{-7}\leq R<4\times10^{-7}$; and the cyan regions are for $4\times10^{-7}\leq R<\times10^{-6}$.}\label{cvct2}
\includegraphics[scale=0.4]{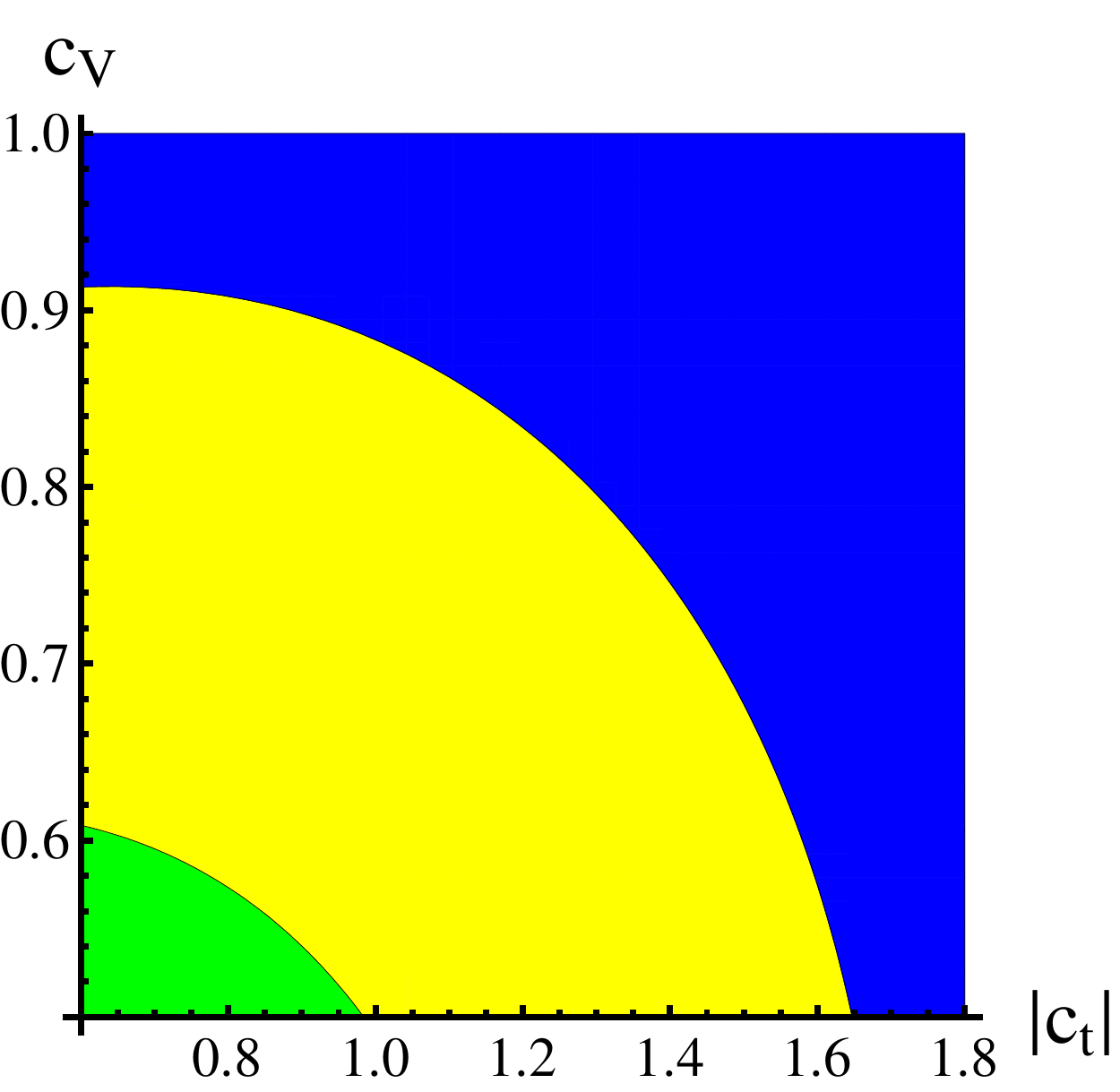}\quad\includegraphics[scale=0.4]{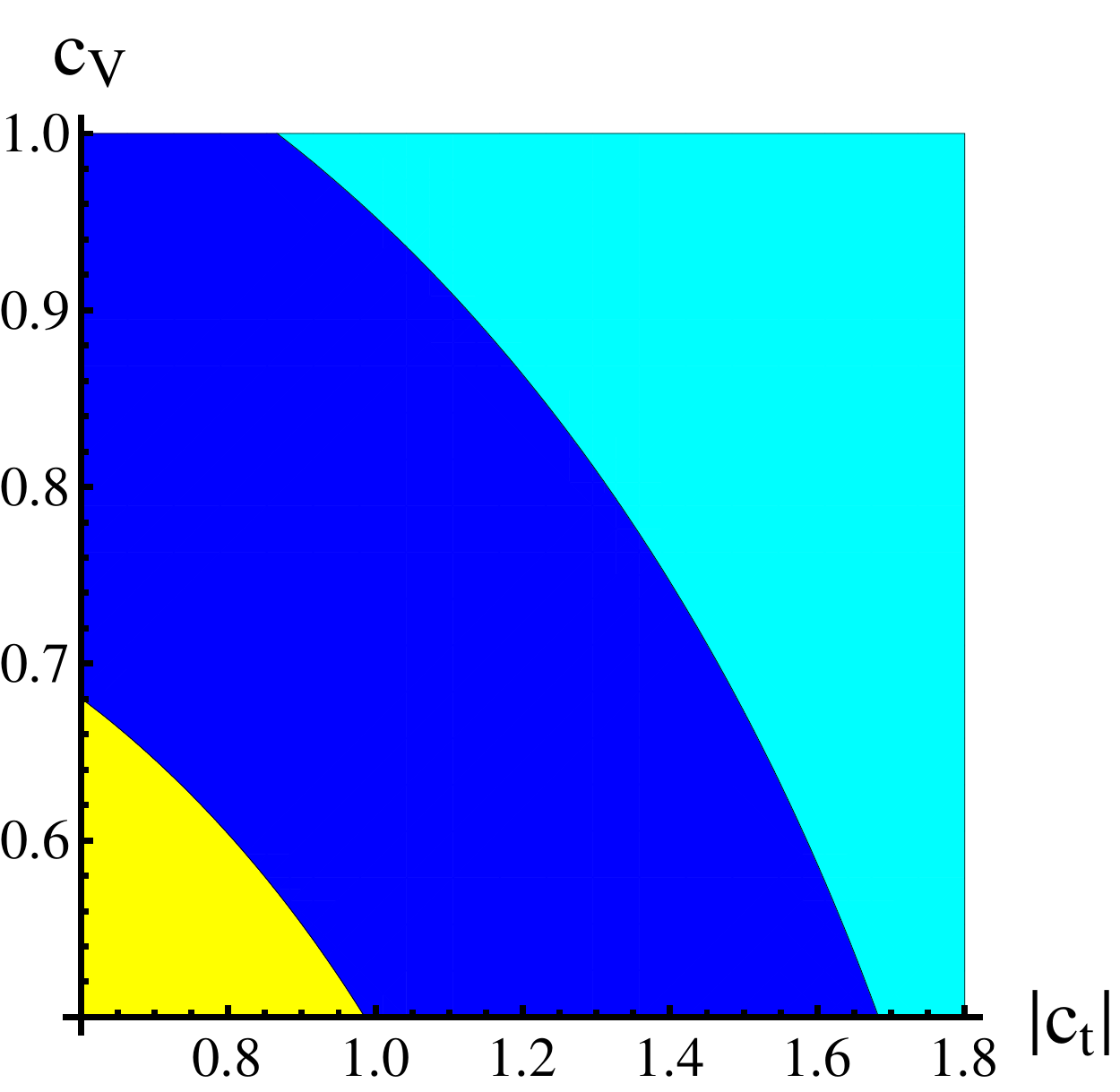}\quad\includegraphics[scale=0.4]{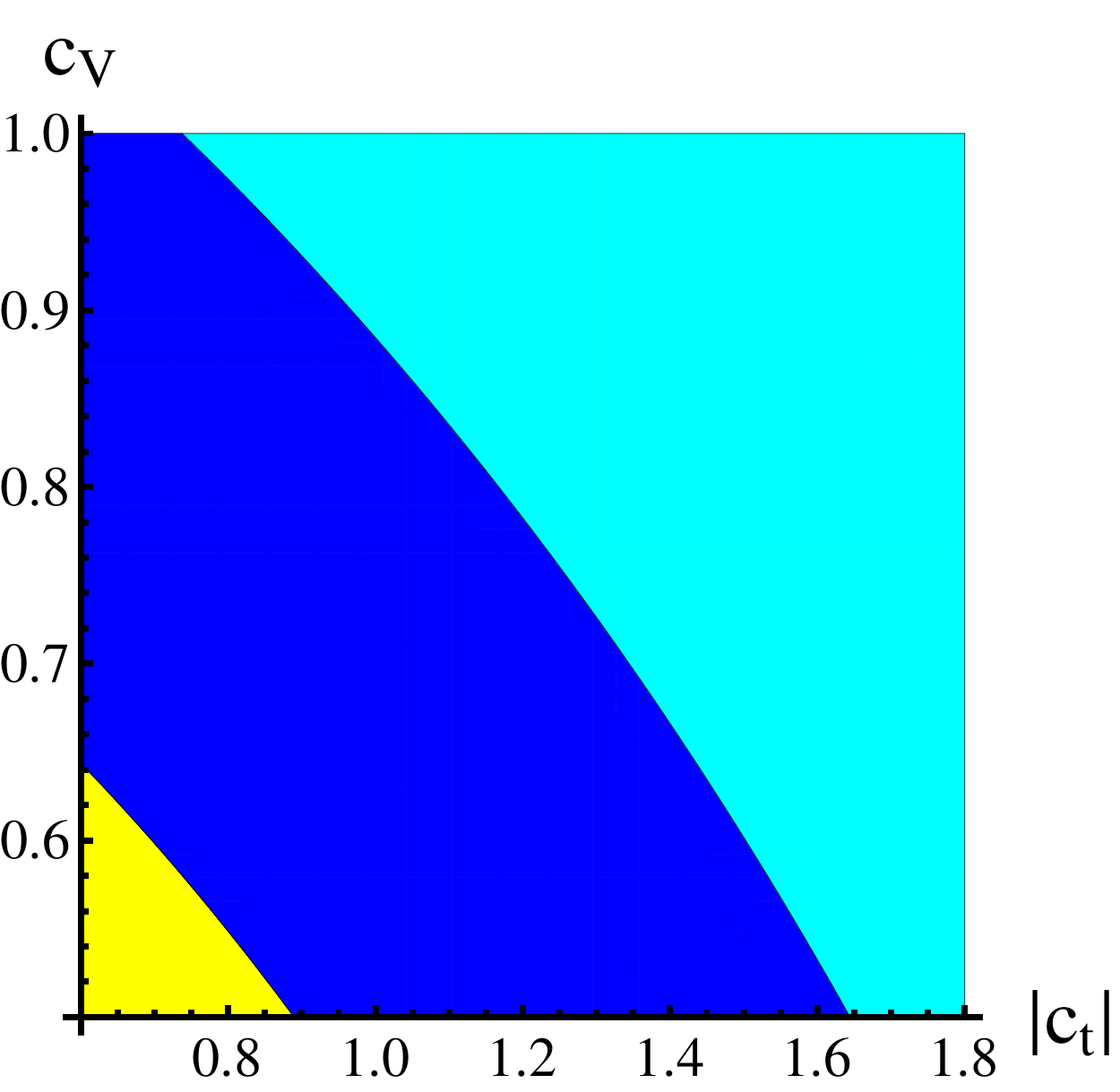}
\end{figure}

According to (\ref{amp}), $R\equiv\textrm{Br}(\tau\rightarrow\mu\gamma)/\textrm{Br}(h\rightarrow\mu\tau)$ is sensitive to the interplay
between $c_V$ and $c_t$. The cancelation between $W$ loop and $t$ loop induced amplitudes would make $R$ very small in some regions
especially for $\alpha_t\sim0$. In \autoref{cvct1} and \autoref{cvct2}, we show some $R\equiv\textrm{Br}(\tau\rightarrow\mu\gamma)/\textrm{Br}(h\rightarrow\mu\tau)$
distribution in $c_V-|c_t|$ plane in unit of $(\Gamma_{h,\textrm{tot}}/\textrm{MeV})$ for some different $\alpha_t$. From the figures,
we can also see the cancelation behavior clearly when $\alpha_t$ is small. For larger $\alpha_t$, the imaginary parts of the amplitudes
would give more important contributions, and the imaginary parts of one loop contribution would also become more important as later
figures shown.

Here and in the following sections, we categorize BSM into two scenarios. In scenario I, we choose most Higgs
couplings close to those in SM, especially $c_V\sim1$ and $\Gamma_h/\Gamma_{h,\textrm{SM}}$. Since the experimental
data \cite{pro} are consistent with the SM predictions, this scenario is popular. While the data still allow the Higgs
couplings away from those in SM and these scenarios are attractive, because they are strongly related to BSM physics.
In scenario II, we choose Lee model \cite{Lee,our} as such a benchmark model. Our previous work \cite{our} showed that
there is no SM limit for the lightest scalar in Lee model. We take the 125 GeV Higgs boson as the lightest one, so
some of its couplings must be away form those in SM, especially $c_V$ should be small. In that paper, we considered full
constraints by data and showed it is still alive. The fitting results for Higgs signal strengths allowed $c_V\sim0.5$,
and at the same time, $|c_b|$ and $\Gamma_h$ must be smaller than those in SM. The results are not sensitive to charged
Higgs loop contribution. The typical $\Gamma_h/\Gamma_{h,\textrm{SM}}\sim\mathcal{O}(0.1)$ for different $|c_b|$ choice.
In both scenarios, $|c_t|\sim|c_{\tau}|\sim1$ are preferred.

\begin{figure}
\caption{$\textrm{Br}(\tau\rightarrow\mu\gamma)$ distributions in $\alpha_t-\alpha_{\tau}$ plane for for
$c_V=\Gamma_h/\Gamma_{h,\textrm{SM}}=1$, taking $|c_t|=0.6,1.2,1.8$ from left to right. The green regions are for
$\textrm{Br}(\tau\rightarrow\mu\gamma)<1.5\times10^{-8}$, the yellow regions are for $1.5\times10^{-8}\leq\textrm{Br}(\tau\rightarrow\mu\gamma)<3.0\times10^{-8}$, and the blue regions are for
$3.0\times10^{-8}\leq\textrm{Br}(\tau\rightarrow\mu\gamma)<4.5\times10^{-8}$. White regions are already excluded
by recent data.}\label{recent1}
\includegraphics[scale=0.4]{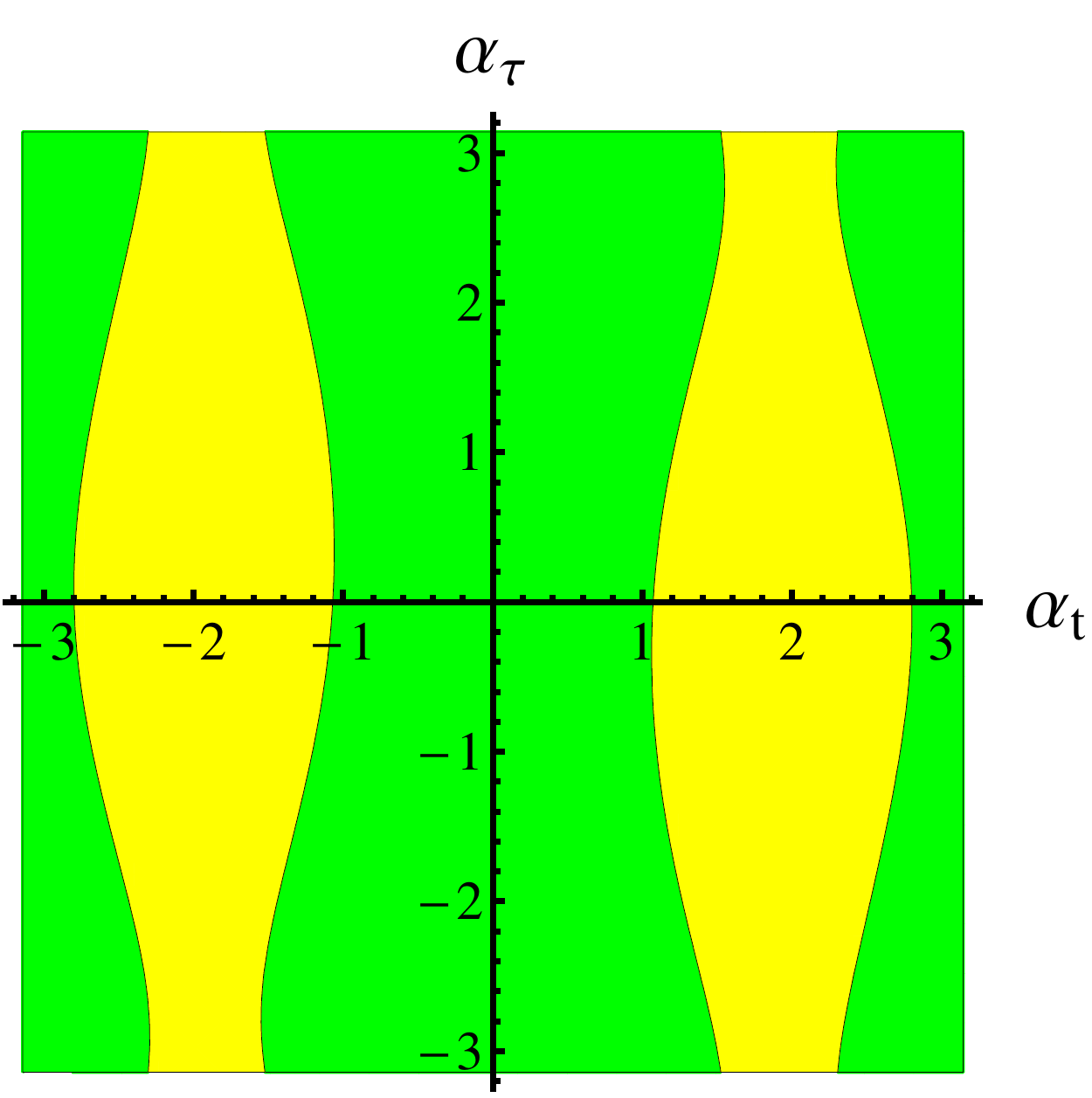}\quad\includegraphics[scale=0.4]{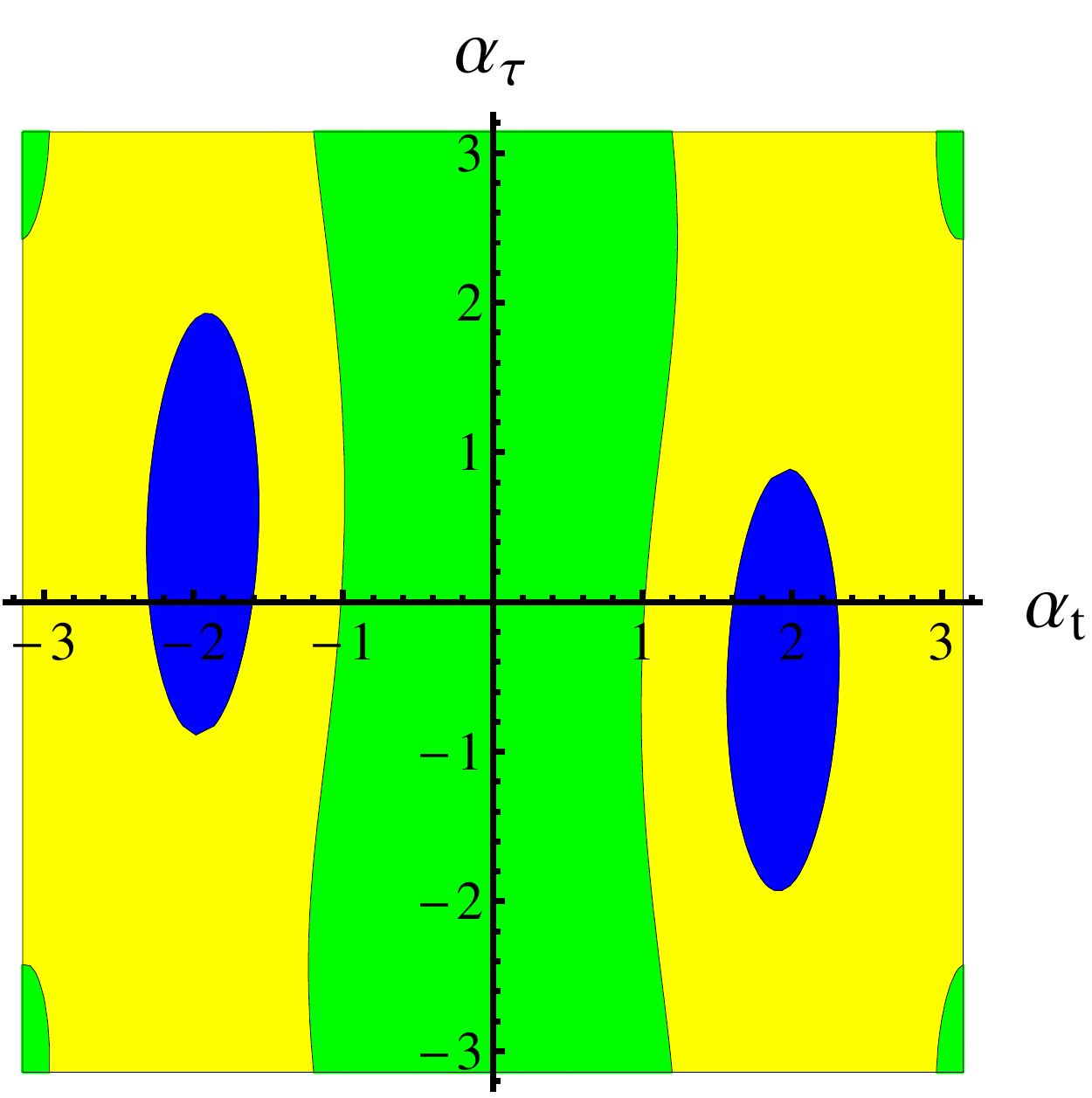}\quad\includegraphics[scale=0.4]{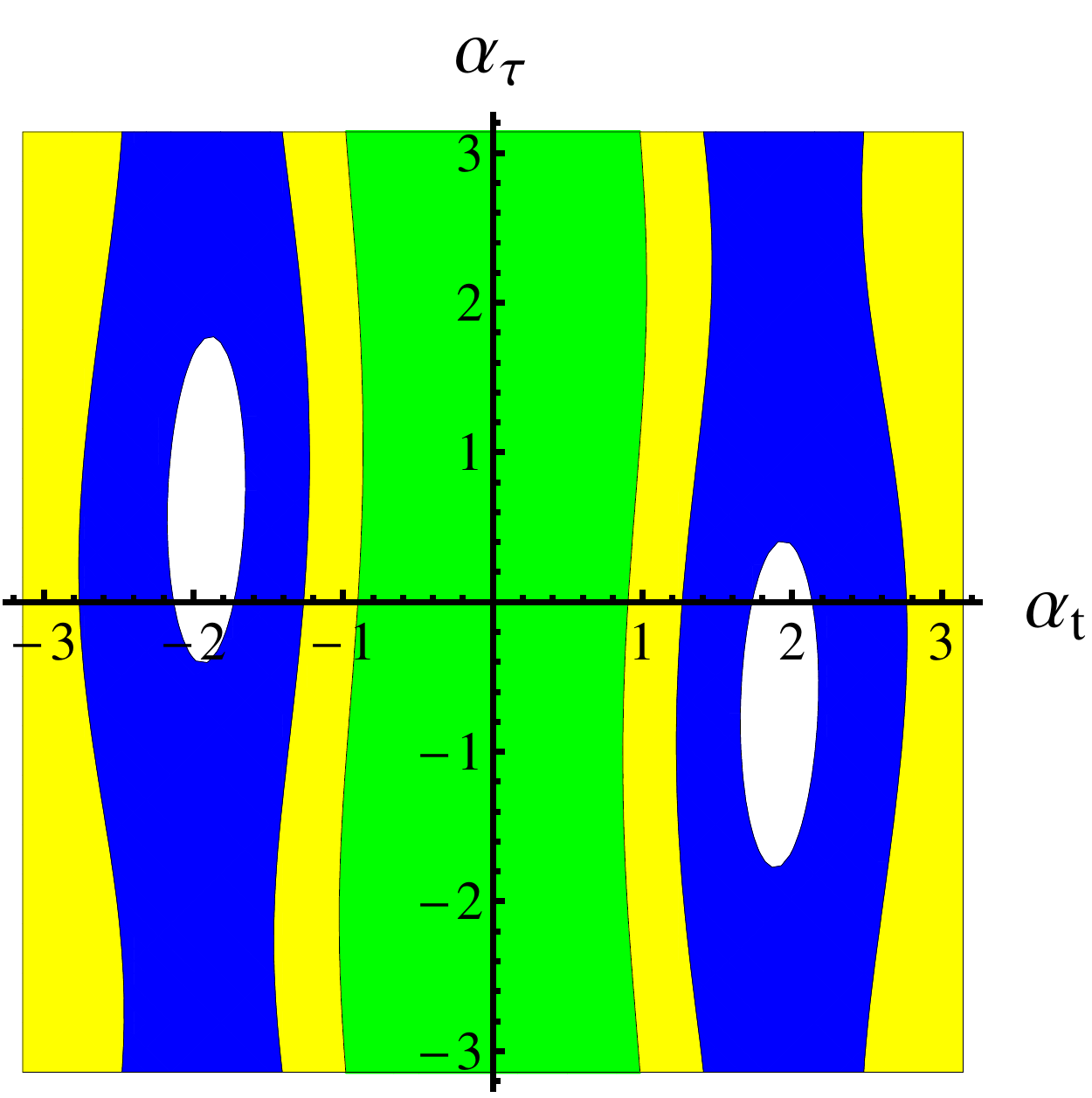}
\end{figure}

\begin{figure}
\caption{$\textrm{Br}(\tau\rightarrow\mu\gamma)$ distributions in $\alpha_t-\alpha_{\tau}$ plane for
$c_V=0.5$, $\Gamma_h/\Gamma_{h,\textrm{SM}}=0.3$, taking $|c_t|=0.6,1.2,1.8$ from left to right.
The green regions are for $\textrm{Br}(\tau\rightarrow\mu\gamma)<2.5\times10^{-9}$, the yellow regions are for $2.5\times10^{-9}\leq\textrm{Br}(\tau\rightarrow\mu\gamma)<5\times10^{-9}$, the blue regions are for
$5\times10^{-9}\leq\textrm{Br}(\tau\rightarrow\mu\gamma)<7.5\times10^{-9}$, and the cyan regions are
for $7.5\times10^{-9}\leq\textrm{Br}(\tau\rightarrow\mu\gamma)<1\times10^{-8}$.}\label{recent2}
\includegraphics[scale=0.4]{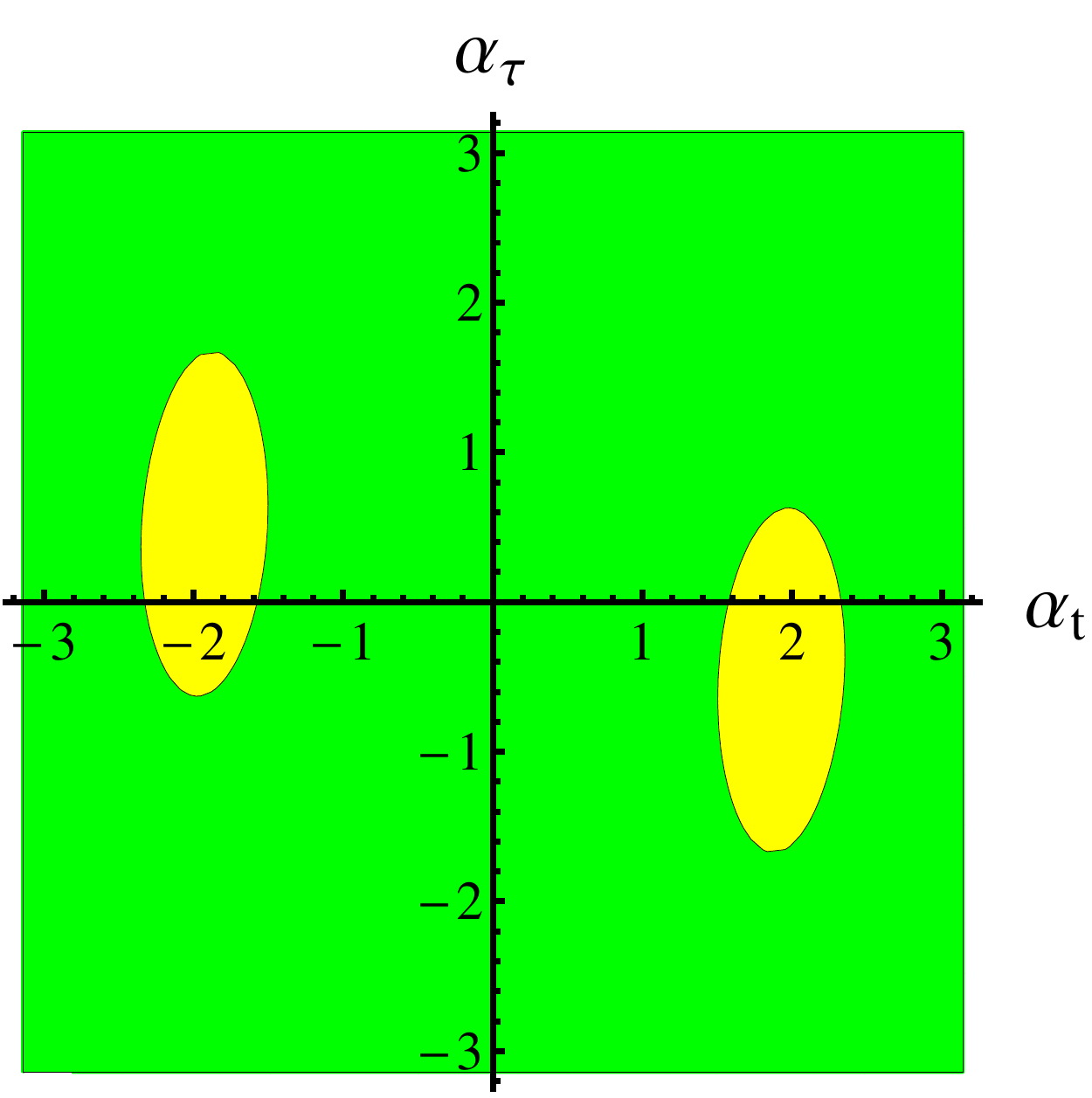}\quad\includegraphics[scale=0.4]{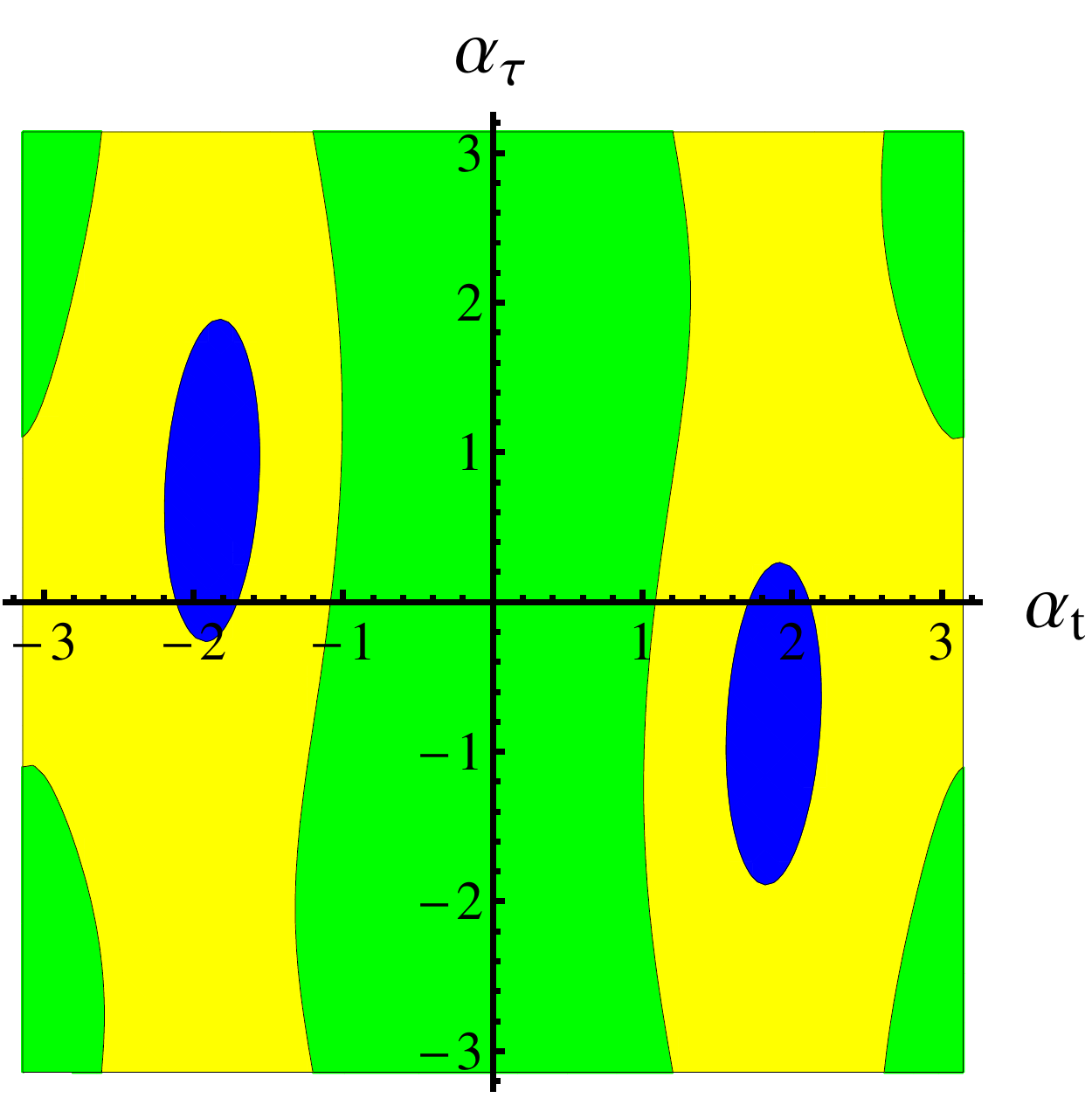}\quad\includegraphics[scale=0.4]{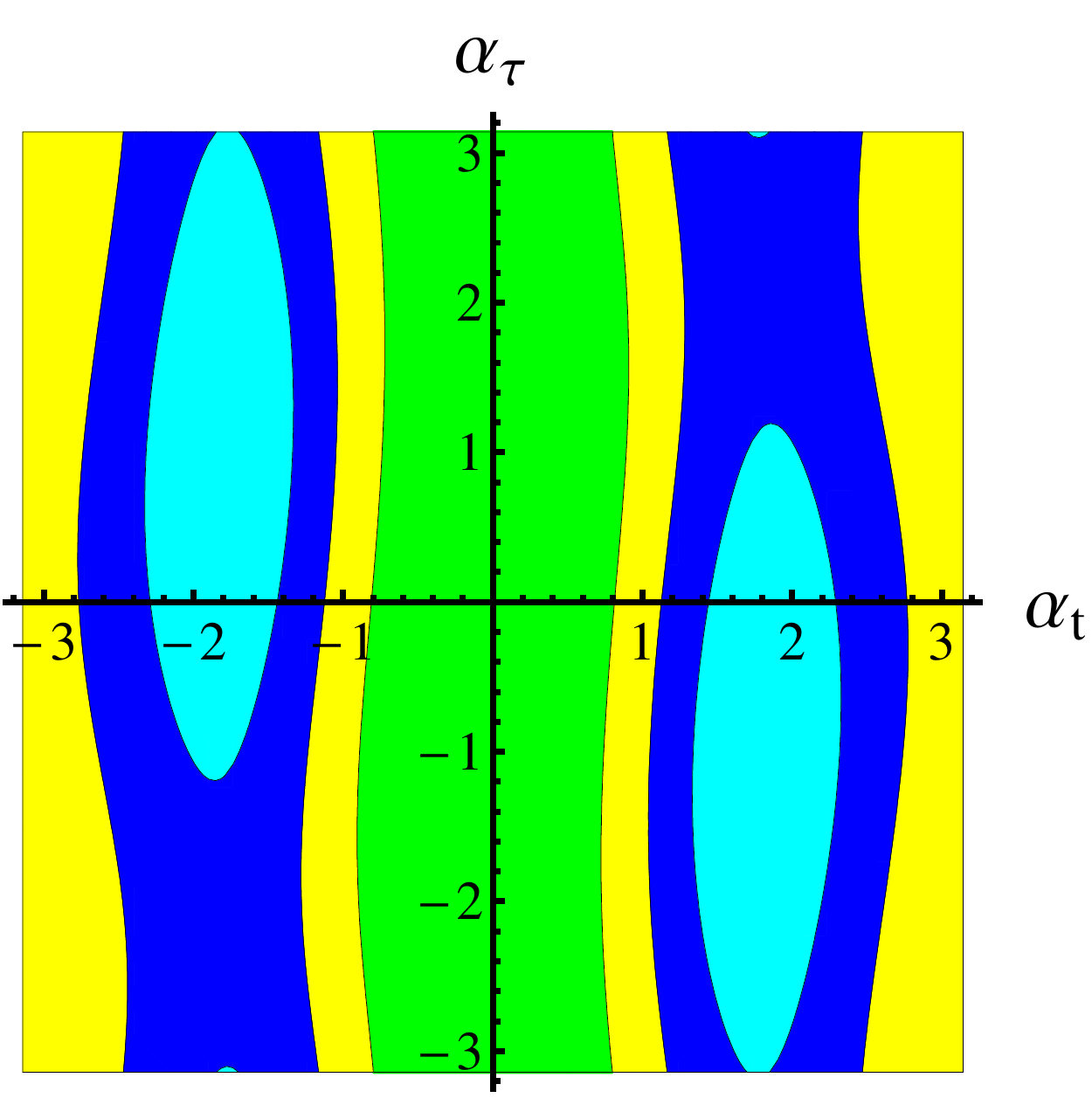}
\end{figure}

In the scenario I, we take $|c_t|=0.6,1.2,1.8$ and plot the predicted branching ratios for $\tau\rightarrow\mu\gamma$
in \autoref{recent1} with $c_V=\Gamma_h/\Gamma_{h,\textrm{SM}}=1$ assuming $\textrm{Br}(h\rightarrow\mu\tau)=1.51\%$
as the CMS upper limit, white regions are already excluded by recent data.
For $|c_t|<1.7$, all the choices for $(\alpha_t, \alpha_{\tau})$ are still allowed by recent data using
this set of benchmark point, thus the recent $\tau\rightarrow\mu\gamma$ measurements cannot give further
constraints. While in the scenario II, the predicted branching ratios for $\tau\rightarrow\mu\gamma$
are highly suppressed to be of $\mathcal{O}(10^{-9})$ that Lee model is not constrained by recent data. We take
$|c_t|=0.6,1.2,1.8$ again and plot the predicted $\textrm{Br}(\tau\rightarrow\mu\gamma)$ in Lee model in
\autoref{recent2} with $c_V=0.5$ and $\Gamma_h/\Gamma_{h,\textrm{SM}}=0.3$, assuming
$\textrm{Br}(h\rightarrow\mu\tau)=1.51\%$ as the CMS upper limit.

\section{Constraints at Future Colliders}
\label{CAFC}
Kopp and Nardecchia \cite{Kandn} studied the phenomenology of $h\rightarrow\mu\tau$ at future LHC ($\sqrt{s}=13\textrm{TeV}$).
With $300\textrm{fb}^{-1}$
luminosity, their results showed that for $\sigma_h=\sigma_{h,\textrm{SM}}$, if no signal is observed, the expected upper limit
at $95\%$ C.L. should be set as $\textrm{Br}(h\rightarrow\mu\tau)<7.7\times10^{-4}$ \cite{Kandn} which means
\begin{equation}
\sqrt{|Y_{\mu\tau}|^2+|Y_{\tau\mu}|^2}<1.1\times10^{-3}\quad\textrm{or}\quad
\sqrt{\frac{(|Y_{\mu\tau}|^2+|Y_{\tau\mu}|^2)v^2}{2m_{\mu}m_{\tau}}}<0.45.
\end{equation}
On the other hand, a signal would be observed at over $3\sigma$ if $\textrm{Br}(h\rightarrow\mu\tau)>1.3\times10^{-3}$ which means
\begin{equation}
\sqrt{|Y_{\mu\tau}|^2+|Y_{\tau\mu}|^2}>1.5\times10^{-3}\quad\textrm{or}\quad
\sqrt{\frac{(|Y_{\mu\tau}|^2+|Y_{\tau\mu}|^2)v^2}{2m_{\mu}m_{\tau}}}>0.59.
\end{equation}
The SuperB factory is a $e^+e^-$ collider at $\Upsilon(4S)$ threshold with the luminosity $75\textrm{ab}^{-1}$.
For the LFV decay $\tau\rightarrow\mu\gamma$, if no signal was observed at the SuperB factory, the expected upper limit at $90\%$ C.L. should be
set as \cite{fut3}
\begin{equation}
\textrm{Br}(\tau\rightarrow\mu\gamma)<2.4\times10^{-9}.
\end{equation}
On the other hand, a signal would be observed at over $3\sigma$ if
\begin{equation}
\textrm{Br}(\tau\rightarrow\mu\gamma)>5.4\times10^{-9}.
\end{equation}
At the Super $\tau$-charm factory, which is a $e^+e^-$ collider at $\sqrt{s}=(2-7)\textrm{GeV}$ with the luminosity
$10\textrm{ab}^{-1}$, there would be about $2.5\times10^{10}$ pairs of $\tau^+\tau^-$ \cite{STC2}. And the sensitivity
for LFV decay $\tau\rightarrow\mu\gamma$ would be of $\mathcal{O}(10^{-10})$ \cite{STC2} because of the suppression
in background compared with that at SuperB factory\footnote{The dominant backgrounds come from $\tau^+\tau^-\gamma$
events with a hard enough photon at SuperB factory; while at Super $\tau$-charm factory, $\sqrt{s}$ is not far away
above the $\tau^+\tau^-$ threshold that almost all photons from $\tau^+\tau^-\gamma$ are soft.}. And the same sensitivity
$(\sim\mathcal{O}(10^{-10}))$ would be also achieved at new Z-factory \cite{zfactory} with $\mathcal{O}(10^{12})$ Z bosons.

For the $\tau\rightarrow\mu\gamma$ results, there are three typical cases listed in \autoref{futurecase}
in which a positive result means a over $3\sigma$ evidence and a negative result means an
exclusion at $90\%$ C.L. as usual. The typical choices are $\textrm{Br}(\tau\rightarrow\mu\gamma)
\sim10^{-8},10^{-9},10^{-10}$ for each case.
\begin{table}[h]
\caption{Choices for typical $\textrm{Br}(\tau\rightarrow\mu\gamma)$ in different cases.}\label{futurecase}
\begin{tabular}{|c|c|c|c|}
\hline
& \begin{tabular}{c}Result at\\ SuperB\end{tabular} & \begin{tabular}{c}Result at\\ Super $\tau$-charm\end{tabular} & \begin{tabular}{c}
Typical choice \\ on $\textrm{Br}(\tau\rightarrow\mu\gamma)$\end{tabular}\\
\hline
Case I & Positive & Positive & $\sim10^{-8}$ \\
\hline
Case II & Negative & Positive & $\sim10^{-9}$ \\
\hline
Case III & Negative & Negative & $\lesssim2\times10^{-10}$ \\
\hline
\end{tabular}
\end{table}
And for $h\rightarrow\mu\tau$ results, we should consider the cases for LHC with positive or negative result separately.

\subsection{LHC with Positive Result}
A positive result in the $h\rightarrow\mu\tau$ search would mean a direct evidence on LFV Higgs-$\mu$-$\tau$ coupling.
We take $(\sigma_h/\sigma_{h,\textrm{SM}})\textrm{Br}(h\rightarrow\mu\tau)=1.5\times10^{-3}$, $3\times10^{-3}$, and
$6\times10^{-3}$ as benchmark points in this subsection.

\begin{figure}
\caption{$\textrm{Br}(\tau\rightarrow\mu\gamma)$ distributions in $\alpha_t-\alpha_{\tau}$ plane for for
$c_V=\Gamma_h/\Gamma_{h,\textrm{SM}}=1$, taking $|c_t|=1$ in the first line and $|c_t|=1.5$ in the second line,
and $(\sigma_h/\sigma_{h,\textrm{SM}})\textrm{Br}(h\rightarrow\mu\tau)=(1.5,3,6)\times10^{-3}$ from left to
right. The green regions are for $\textrm{Br}(\tau\rightarrow\mu\gamma)<2.4\times10^{-9}$, the yellow regions are for $2.4\times10^{-9}\leq\textrm{Br}(\tau\rightarrow\mu\gamma)<5.4\times10^{-9}$, the blue regions are for
$5.4\times10^{-9}\leq\textrm{Br}(\tau\rightarrow\mu\gamma)<1\times10^{-8}$, and the cyan regions are
for $\textrm{Br}(\tau\rightarrow\mu\gamma)\geq1\times10^{-8}$.}\label{futsm1}
\includegraphics[scale=0.4]{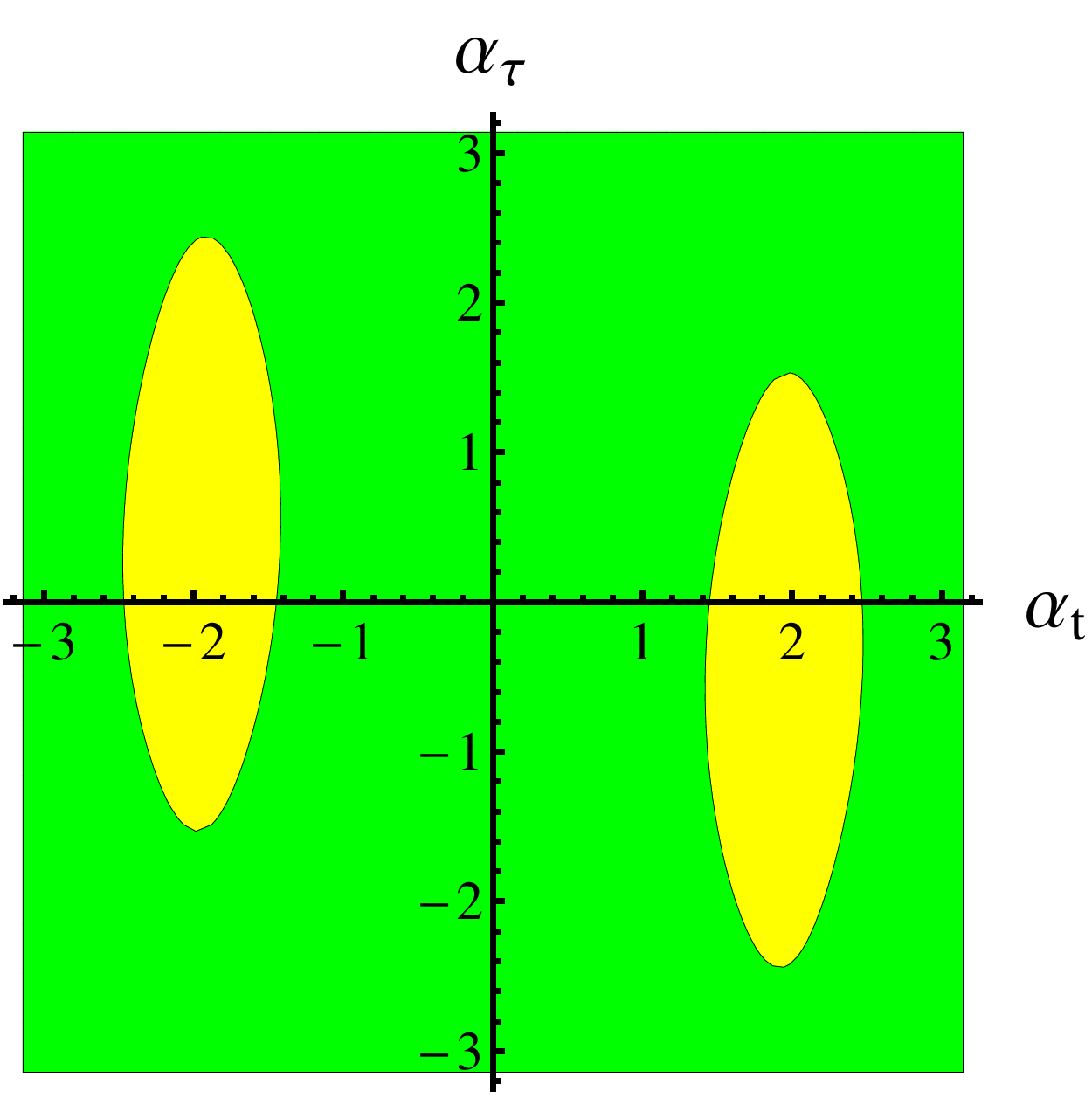}\quad\includegraphics[scale=0.4]{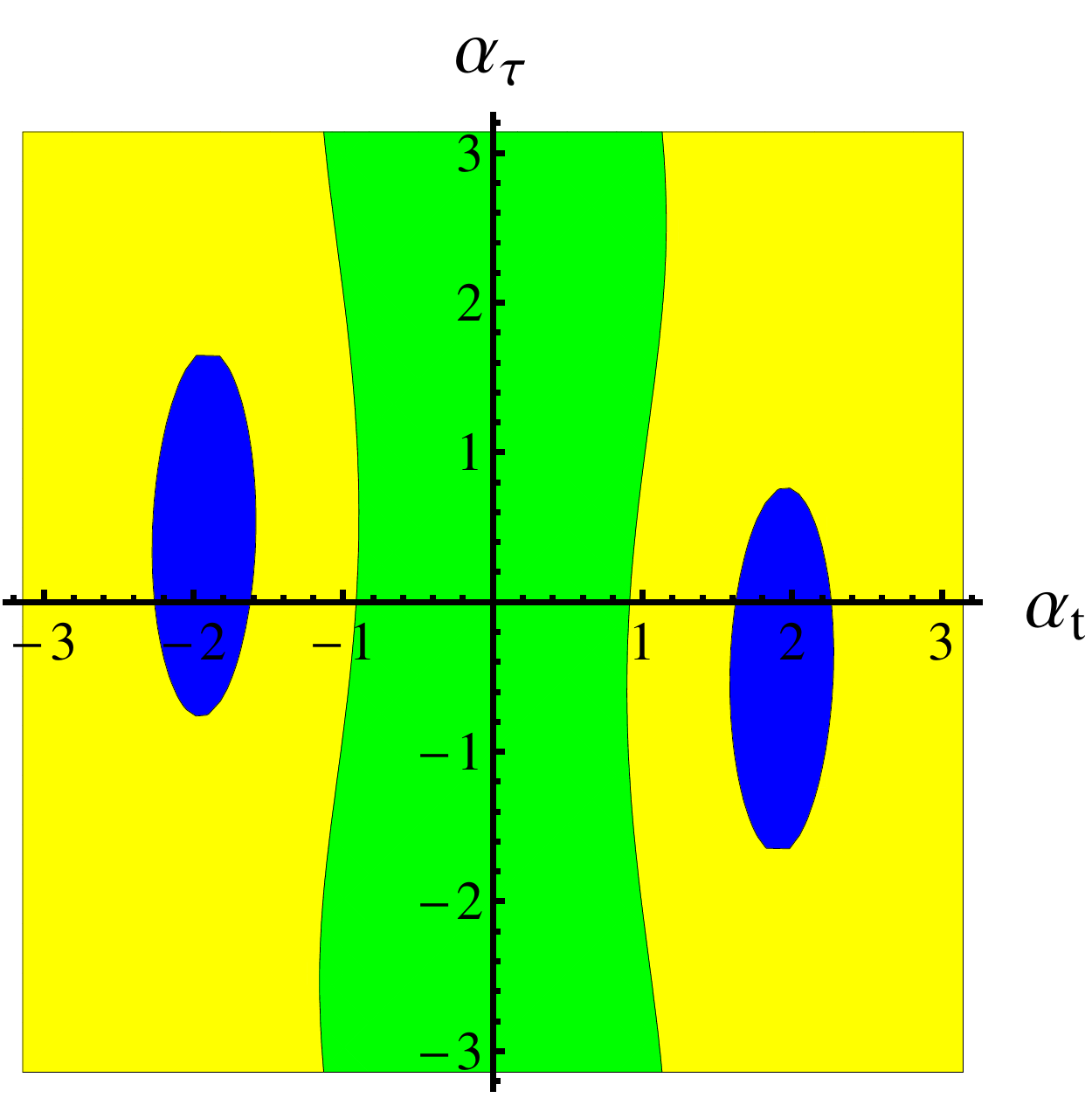}\quad\includegraphics[scale=0.4]{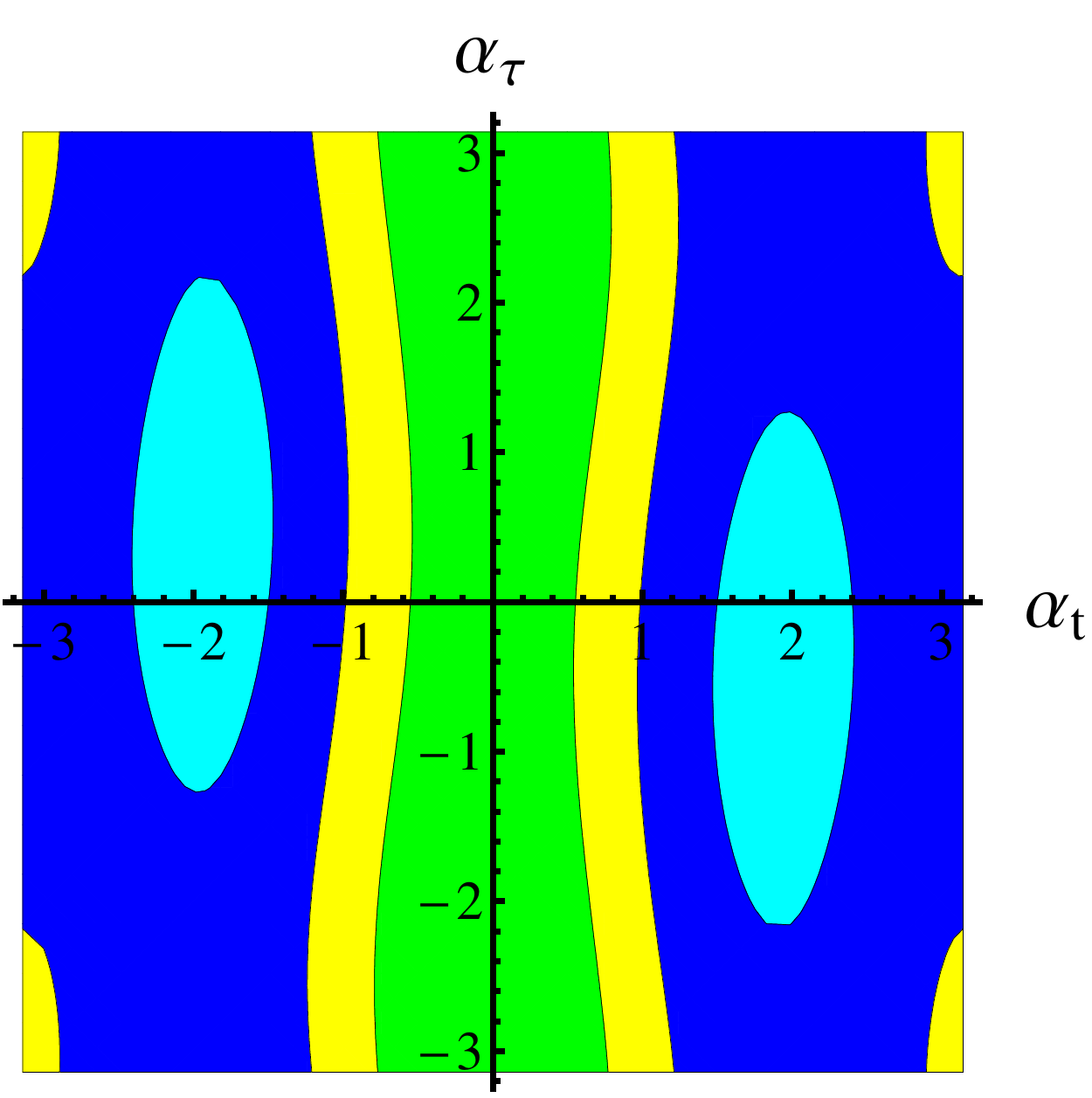}\\
\includegraphics[scale=0.4]{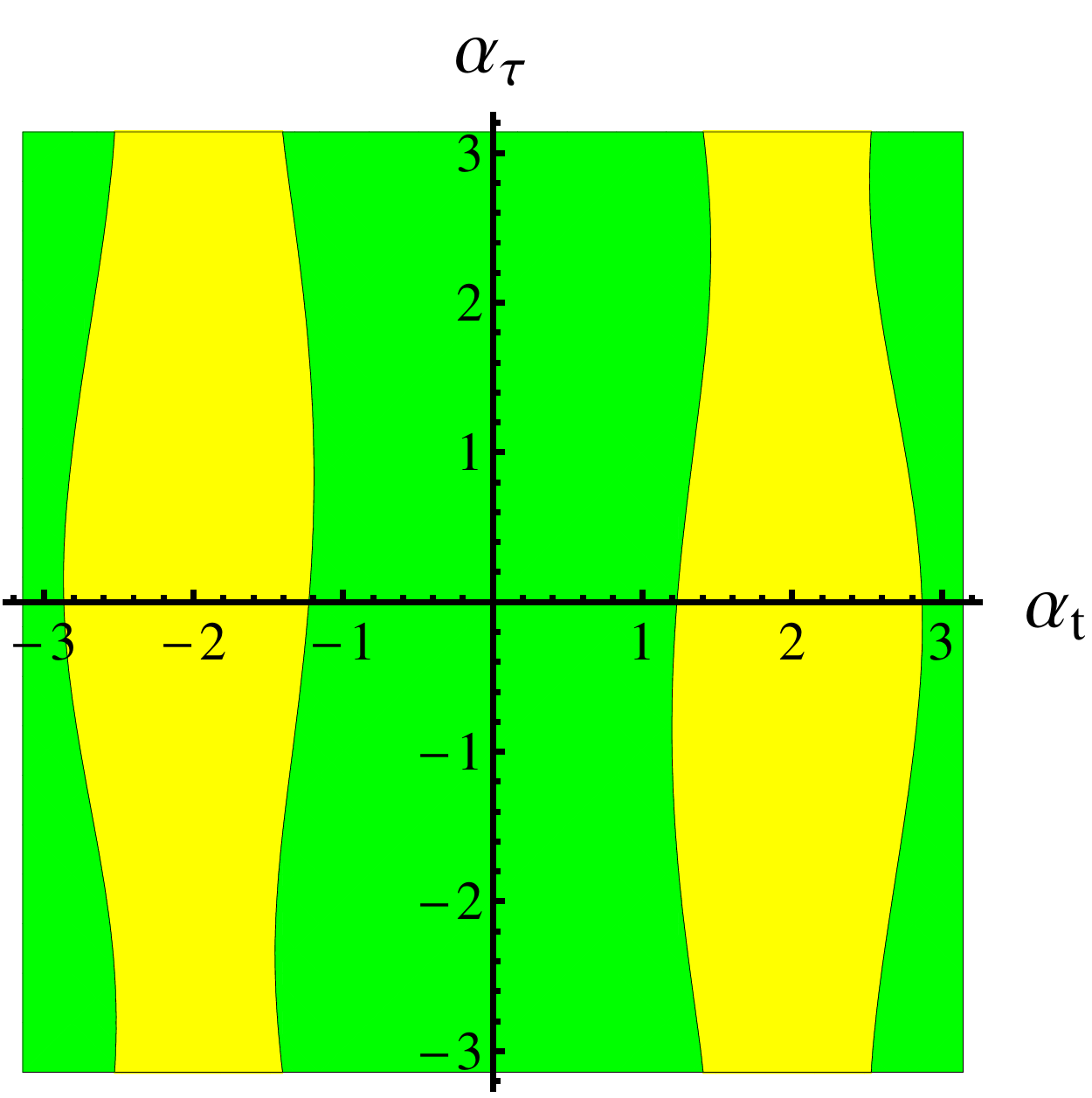}\quad\includegraphics[scale=0.4]{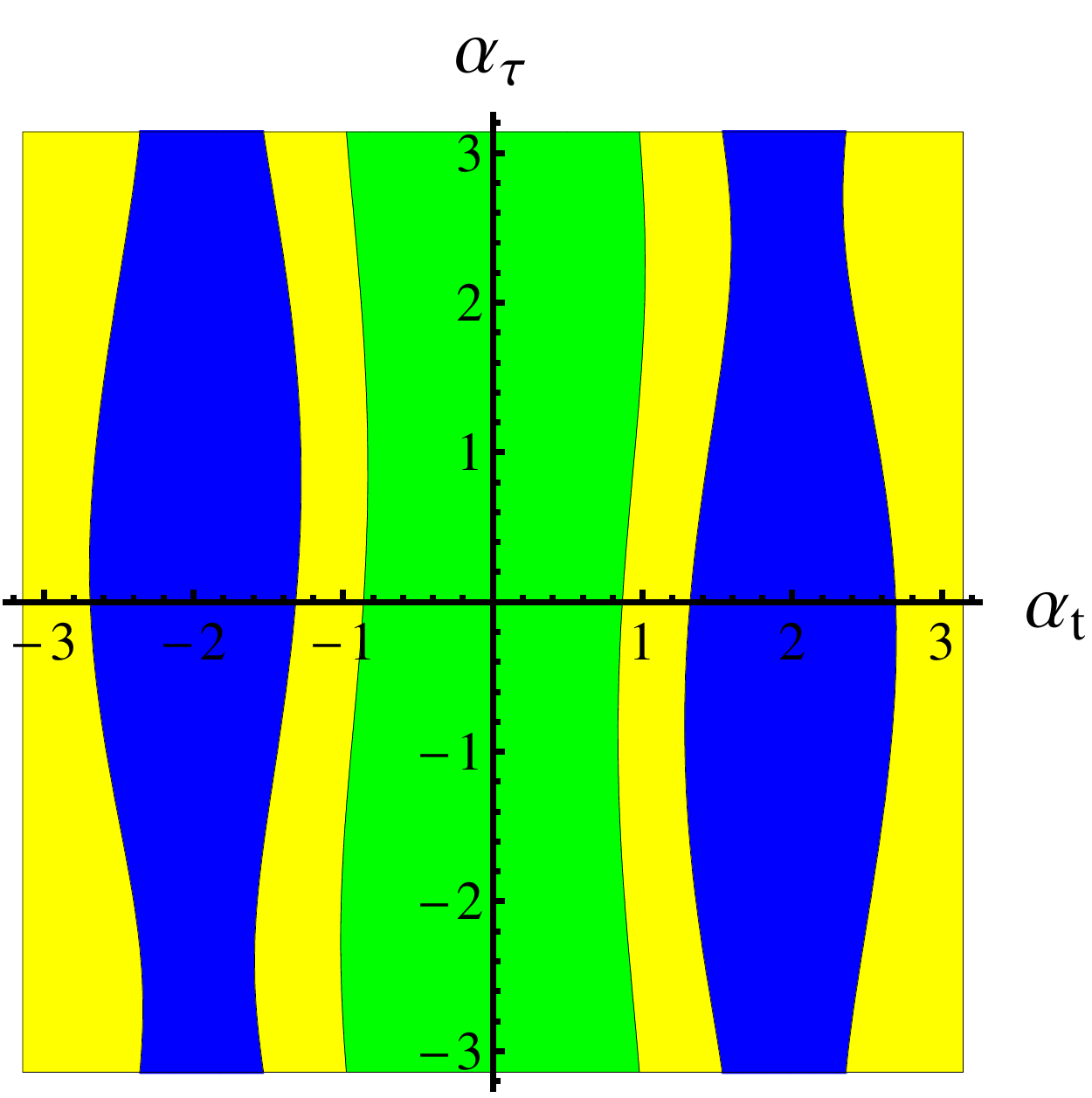}\quad\includegraphics[scale=0.4]{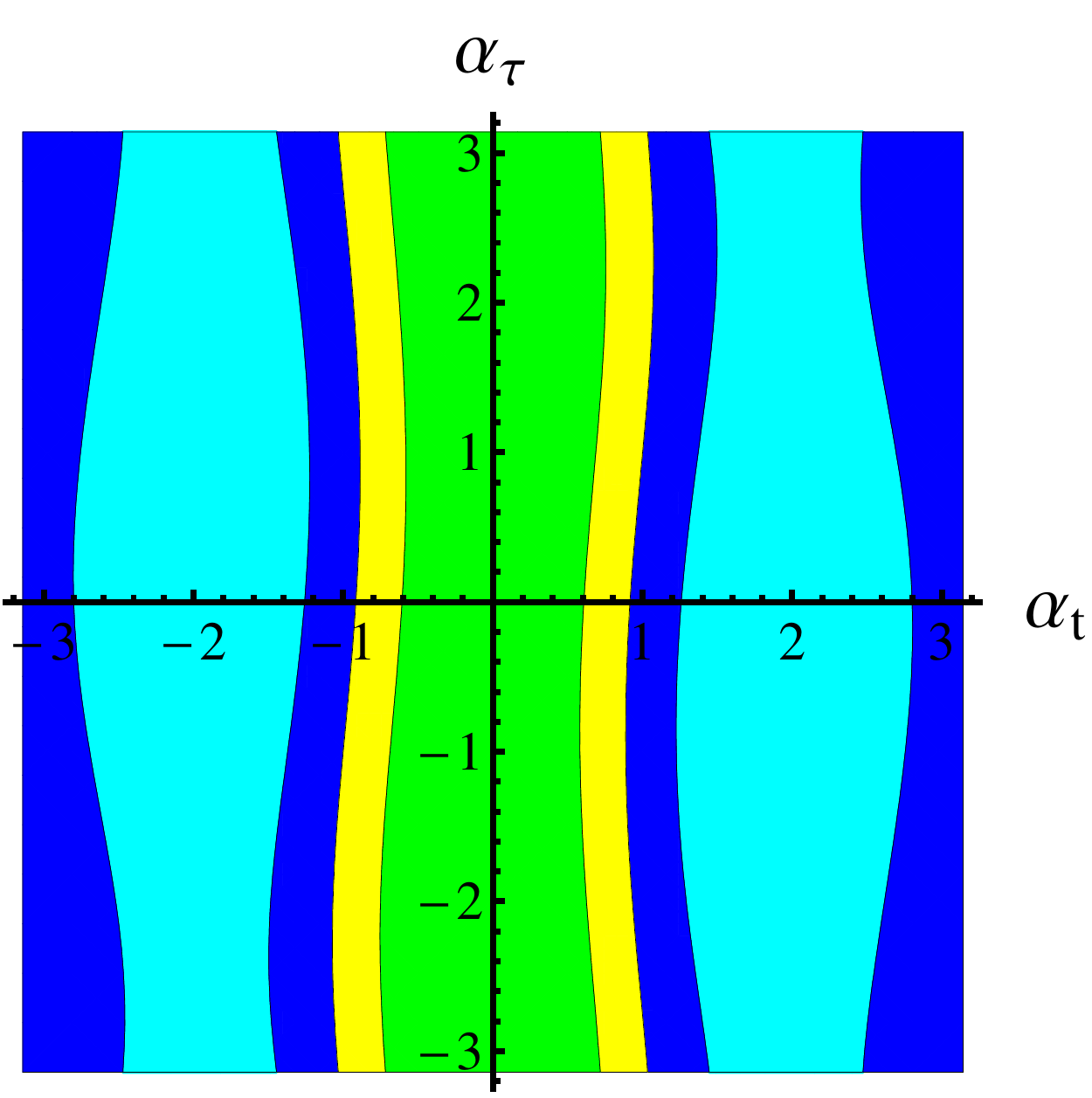}
\end{figure}

First, consider scenario I in \autoref{CBRE} where the coupling strengths are close to those in the SM. Taking
$c_V=\Gamma_h/\Gamma_{h,\textrm{SM}}=1$, $|c_t|=1$ and $1.5$, we show the $\textrm{Br}(\tau\rightarrow\mu\gamma)$
distributions in $\alpha_t-\alpha_{\tau}$ plane in \autoref{futsm1} with the boundaries set according to the
sensitivity of SuperB factory.
We can see that if $(\sigma_h/\sigma_{h,\textrm{SM}})\textrm{Br}(h\rightarrow\mu\tau)\gtrsim(2-3)\times10^{-3}$, the
typical predicted $\textrm{Br}(\tau\rightarrow\mu\gamma)$ would reach the SuperB sensitivity. While if $(\sigma_h/\sigma_{h,\textrm{SM}})\textrm{Br}(h\rightarrow\mu\tau)$
was smaller, the $\tau\rightarrow\mu\gamma$ process would not be found at SuperB factory.

\begin{figure}
\caption{$\textrm{Br}(\tau\rightarrow\mu\gamma)$ distributions in $\alpha_t-\alpha_{\tau}$ plane for for
$c_V=\Gamma_h/\Gamma_{h,\textrm{SM}}=1$, taking $|c_t|=1$ in the first line and $|c_t|=1.5$ in the second line,
and $(\sigma_h/\sigma_{h,\textrm{SM}})\textrm{Br}(h\rightarrow\mu\tau)=(1.5,3,6)\times10^{-3}$ from left to
right. The green regions are for $\textrm{Br}(\tau\rightarrow\mu\gamma)<2\times10^{-10}$, the yellow regions are for $2\times10^{-10}\leq\textrm{Br}(\tau\rightarrow\mu\gamma)<5\times10^{-10}$, the blue regions are for
$5\times10^{-10}\leq\textrm{Br}(\tau\rightarrow\mu\gamma)<10^{-9}$, and the cyan regions are
for $10^{-9}\leq\textrm{Br}(\tau\rightarrow\mu\gamma)<2.4\times10^{-9}$.}\label{futsm2}
\includegraphics[scale=0.4]{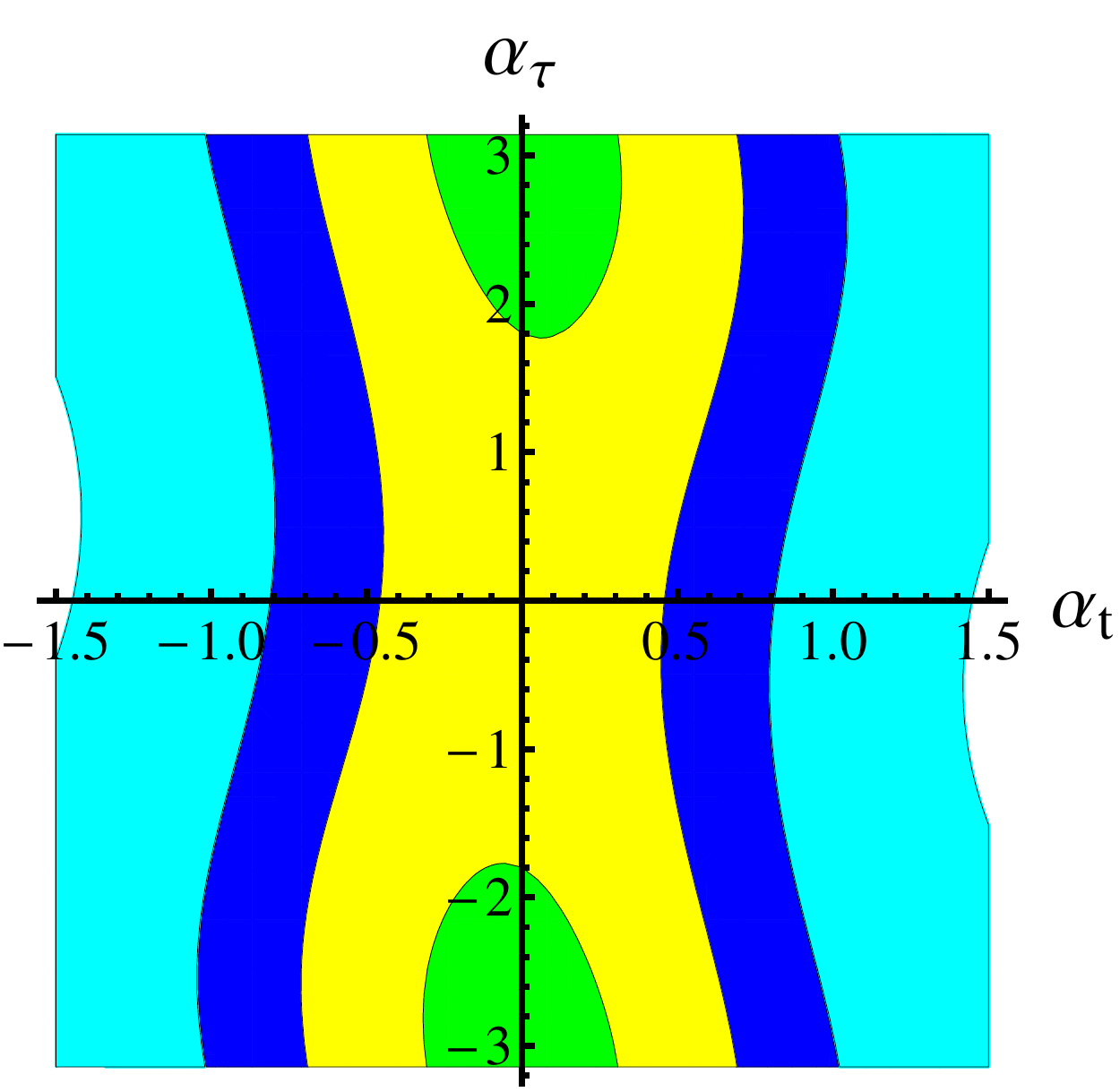}\quad\includegraphics[scale=0.4]{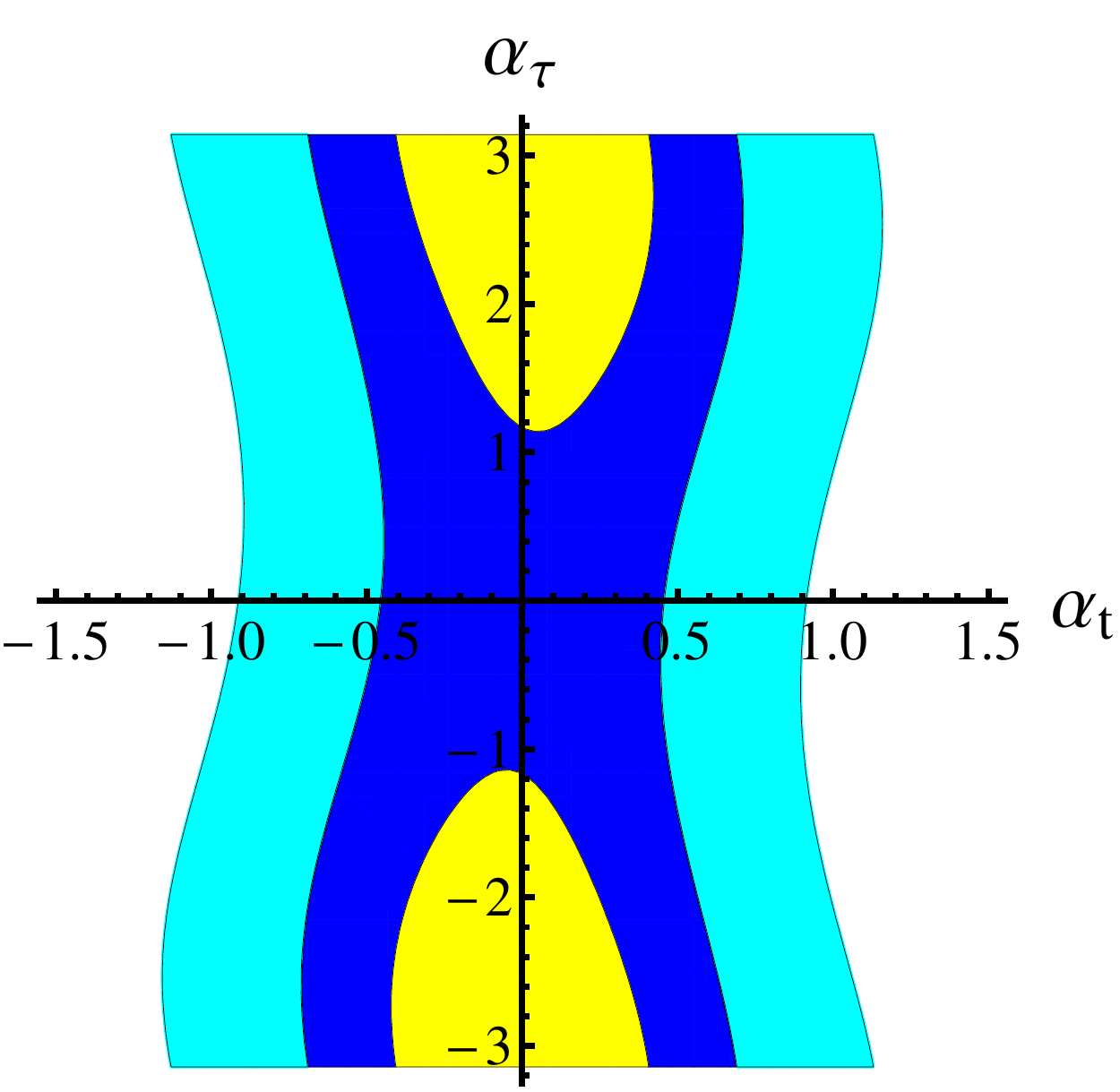}\quad\includegraphics[scale=0.4]{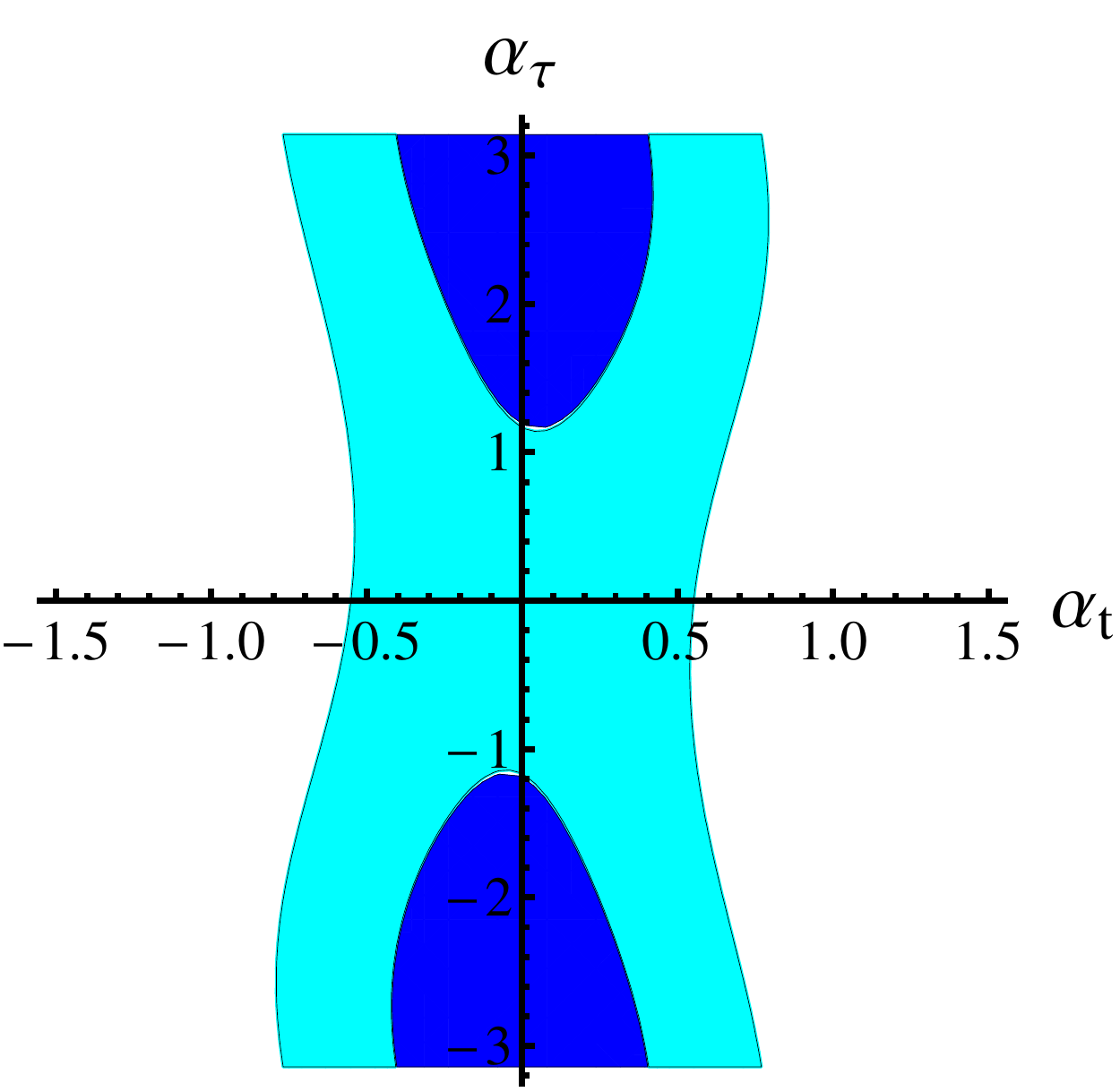}\\
\includegraphics[scale=0.4]{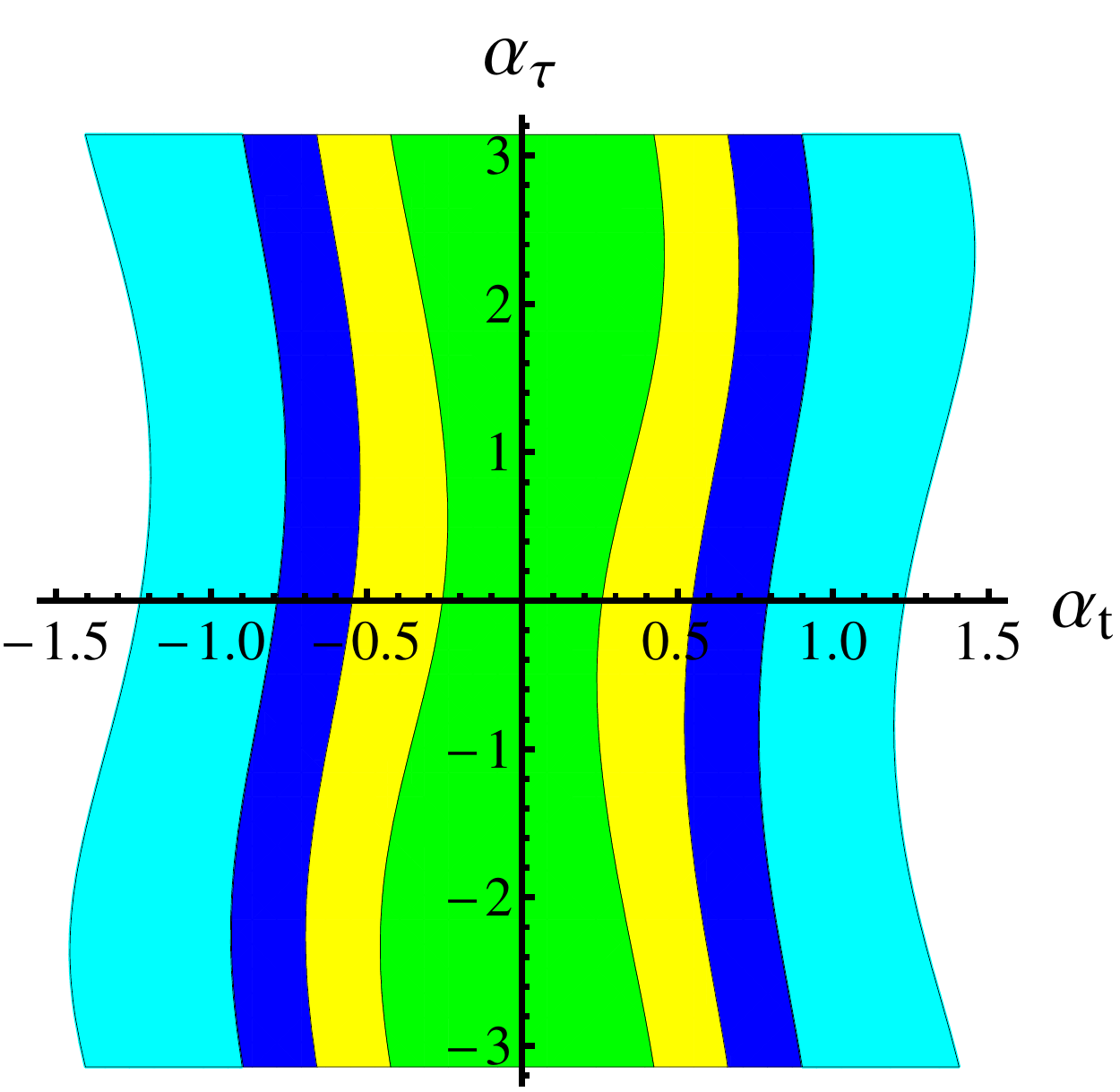}\quad\includegraphics[scale=0.4]{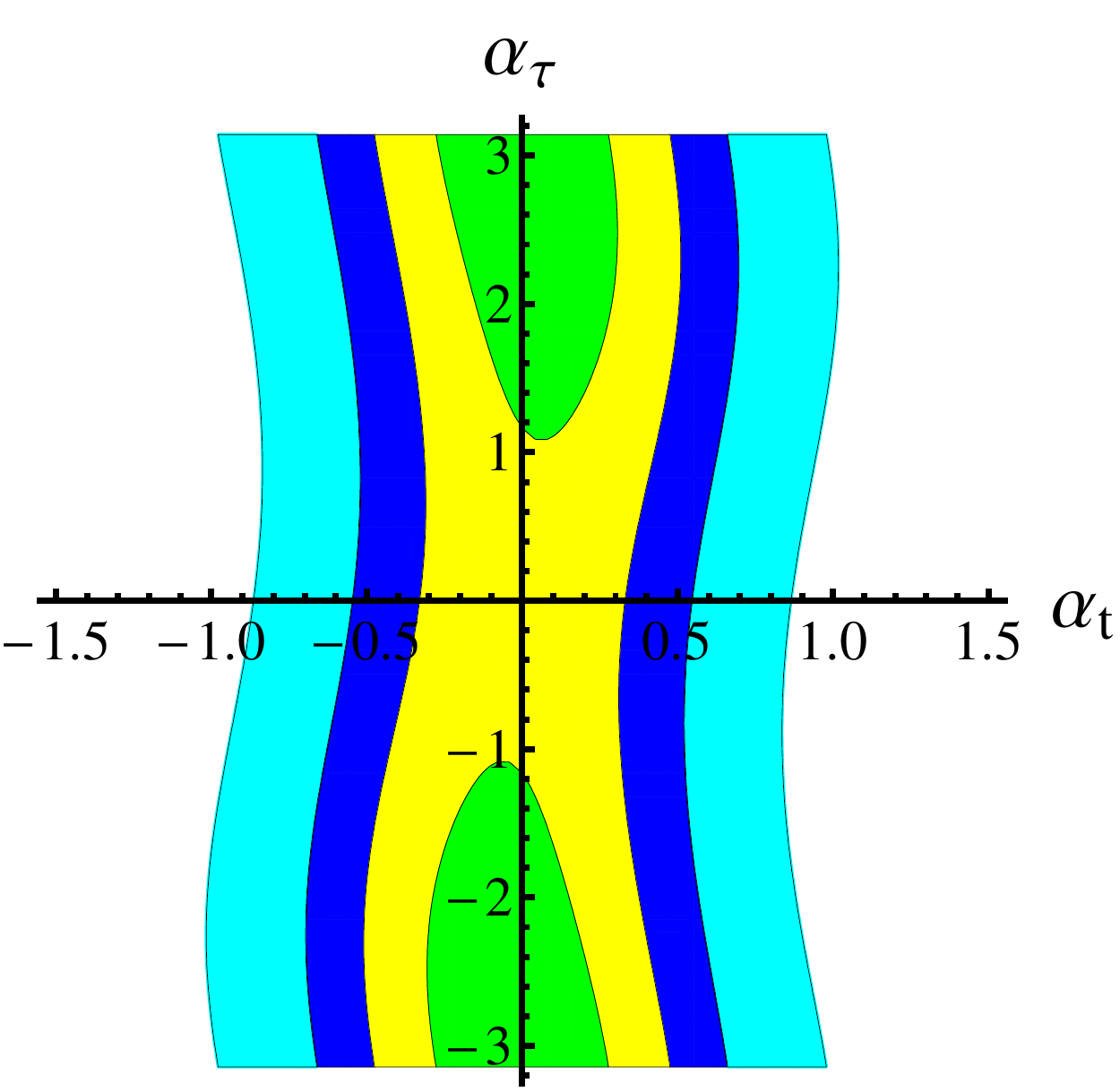}\quad\includegraphics[scale=0.4]{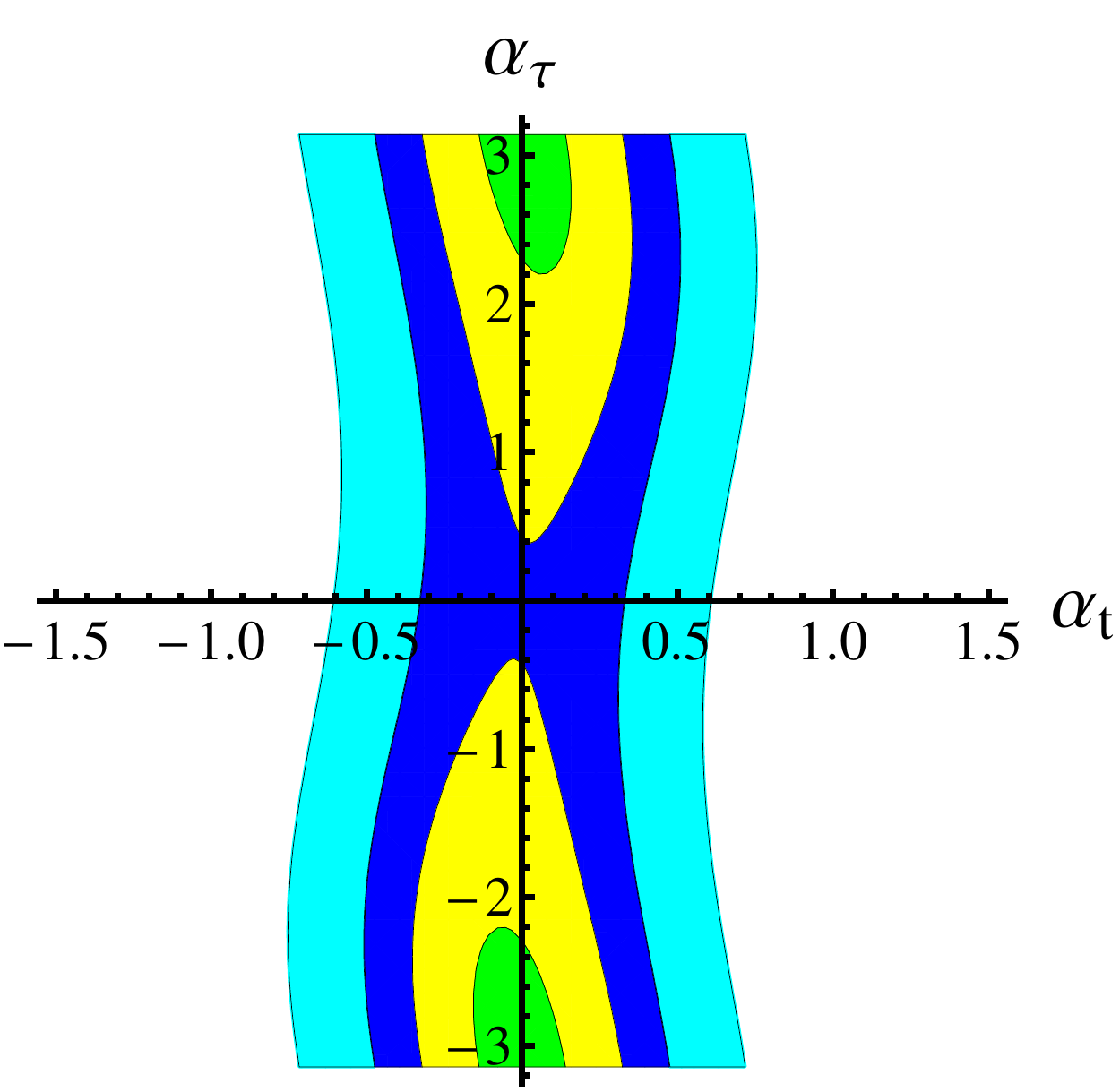}
\end{figure}

Then we should focus on the green regions which mean the cases with negative results at SuperB factory. Here we show
the $\textrm{Br}(\tau\rightarrow\mu\gamma)$ distributions in $\alpha_t-\alpha_{\tau}$ plane in \autoref{futsm2} with
the boundaries set according to the sensitivity of Super $\tau$-charm factory. For $\textrm{Br}(\tau\rightarrow\mu\gamma)
\sim10^{-9}$ or smaller, $|\alpha_t|\lesssim1.5$ were favored.
If LHC gave positive results, the typical predicted $\textrm{Br}(\tau\rightarrow\mu\gamma)$ must reach the sensitivity
of Super $\tau$-charm factory in this scenario. If Super $\tau$-charm factory gave negative results,
it would give strict constraints on the Higgs couplings.

In summary, For case I in \autoref{futurecase}, if the SuperB factory gave positive results in searching
$\tau\rightarrow\mu\gamma$ (thus it must be discovered at Super $\tau$-charm factory as well),
$(\alpha_t,\alpha_{\tau})$ would fall into the blue or cyan regions in \autoref{futsm1}. The value of $\alpha_{\tau}$
was usually free for larger $|c_t|$ and $(\sigma_h/\sigma_{h,\textrm{SM}})\textrm{Br}(h\rightarrow\mu\tau)$, while
$|\alpha_t|\gtrsim1$ were more favored for any case. While for case II in \autoref{futurecase}, SuperB factory gave
negative results but Super $\tau$-charm factory gave positive results, $|\alpha_t|\lesssim1$ would be favored, but for
most cases there would be no constraints on $\alpha_{\tau}$. For case III in \autoref{futurecase} that both factories
gave negative results, larger $|\alpha_{\tau}|$ and $|c_t|$ would be favored.

Second, consider Lee model which is scenario II in \autoref{CBRE}. In this scenario, both $c_V$ and $\Gamma_h/
\Gamma_{h,\textrm{SM}}$ are smaller that the predicted $\textrm{Br}(\tau\rightarrow\mu\gamma)$ are smaller. For
example, taking $(\sigma_h/\sigma_{h,\textrm{SM}})\textrm{Br}(h\rightarrow\mu\tau)=3\times10^{-3}$ as a benchmark
point, the predicted $\textrm{Br}(\tau\rightarrow\mu\gamma)\lesssim(0.8-1.6)\times10^{-9}$ which cannot lead to a
positive result at SuperB factory.

\begin{figure}
\caption{$\textrm{Br}(\tau\rightarrow\mu\gamma)$ distributions in $\alpha_t-\alpha_{\tau}$ plane for for
$c_V=0.5$ and $\Gamma_h/\Gamma_{h,\textrm{SM}}=0.3$, taking $|c_t|=0.8$ in the first line and $|c_t|=1.2$
in the second line, and $(\sigma_h/\sigma_{h,\textrm{SM}})\textrm{Br}(h\rightarrow\mu\tau)=(1.5,3,6)\times10^{-3}$
from left to right. The green regions are for $\textrm{Br}(\tau\rightarrow\mu\gamma)<2\times10^{-10}$, the yellow regions are for $2\times10^{-10}\leq\textrm{Br}(\tau\rightarrow\mu\gamma)<5\times10^{-10}$, the blue regions are for
$5\times10^{-10}\leq\textrm{Br}(\tau\rightarrow\mu\gamma)<10^{-9}$, and the cyan regions are
for $10^{-9}\leq\textrm{Br}(\tau\rightarrow\mu\gamma)<2.4\times10^{-9}$.}\label{futLee}
\includegraphics[scale=0.4]{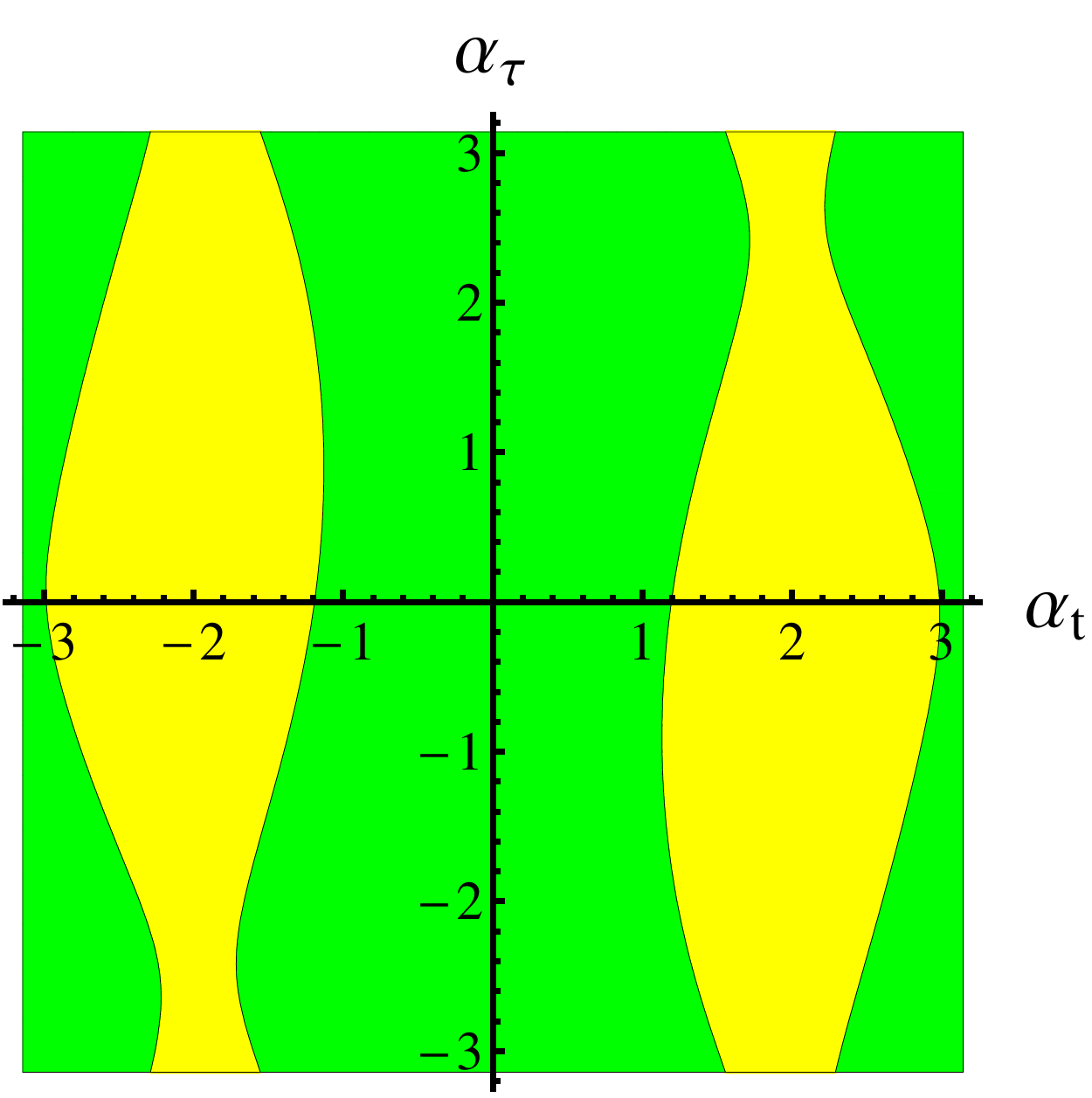}\quad\includegraphics[scale=0.4]{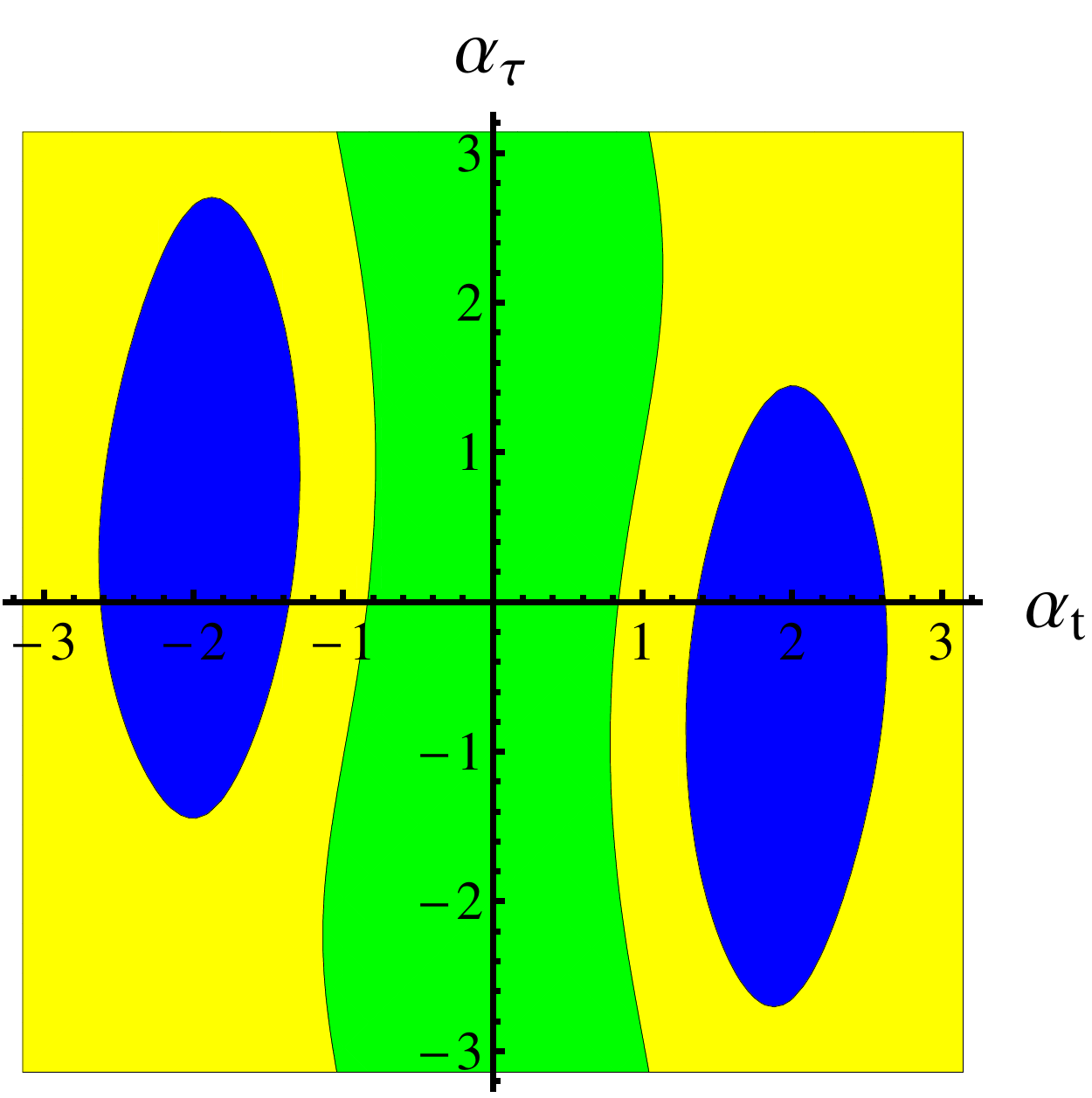}\quad\includegraphics[scale=0.4]{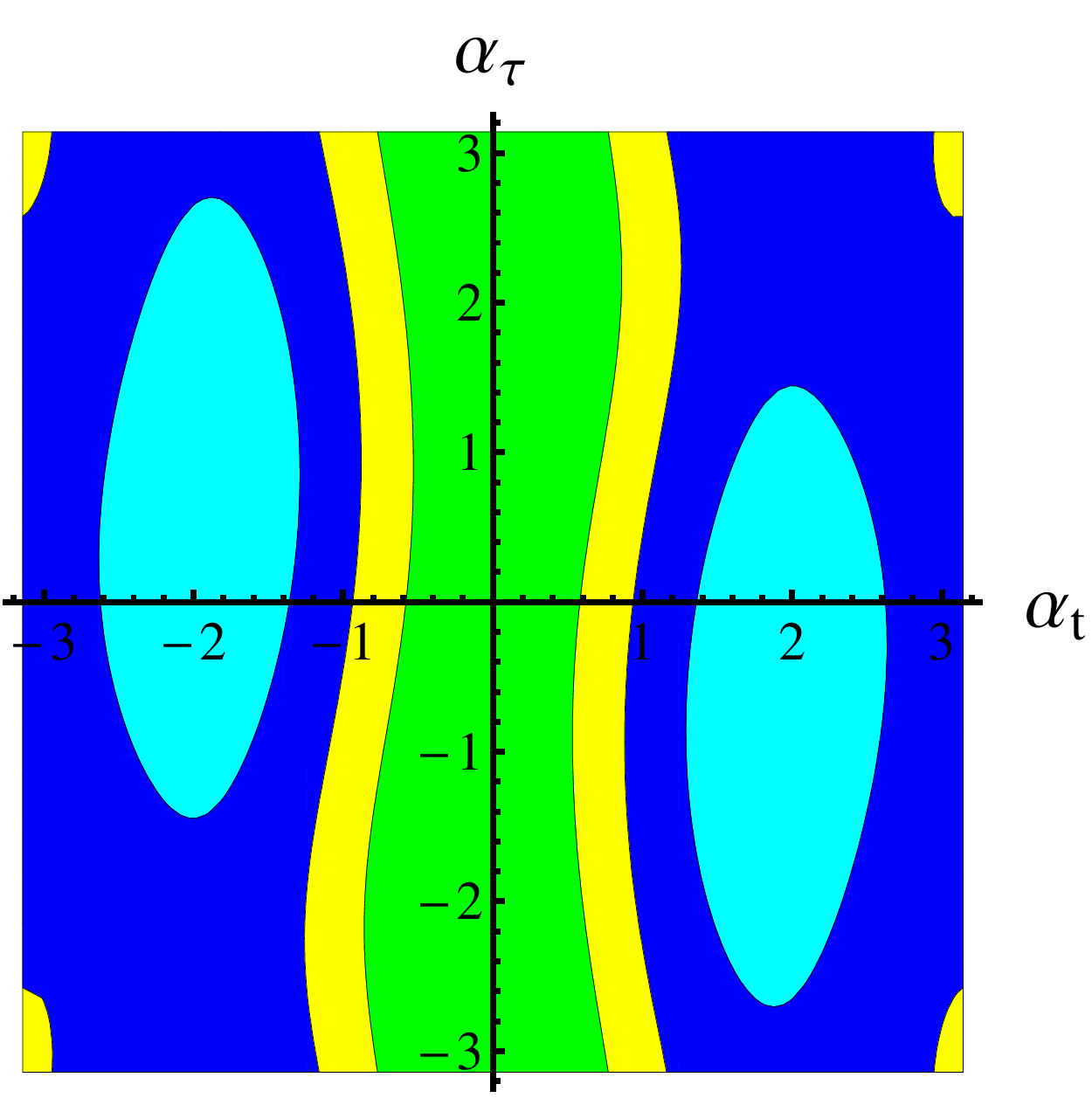}\\
\includegraphics[scale=0.4]{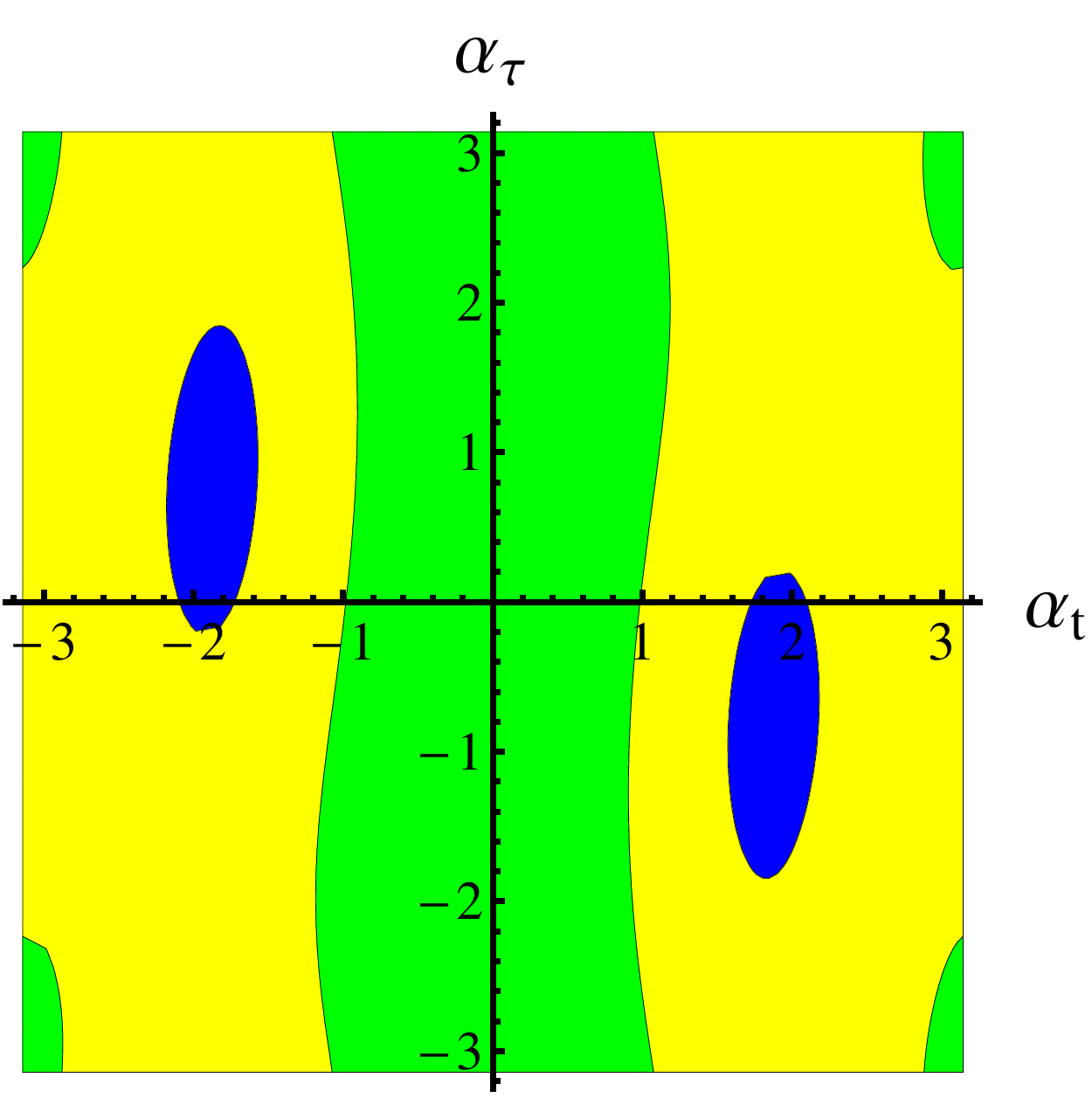}\quad\includegraphics[scale=0.4]{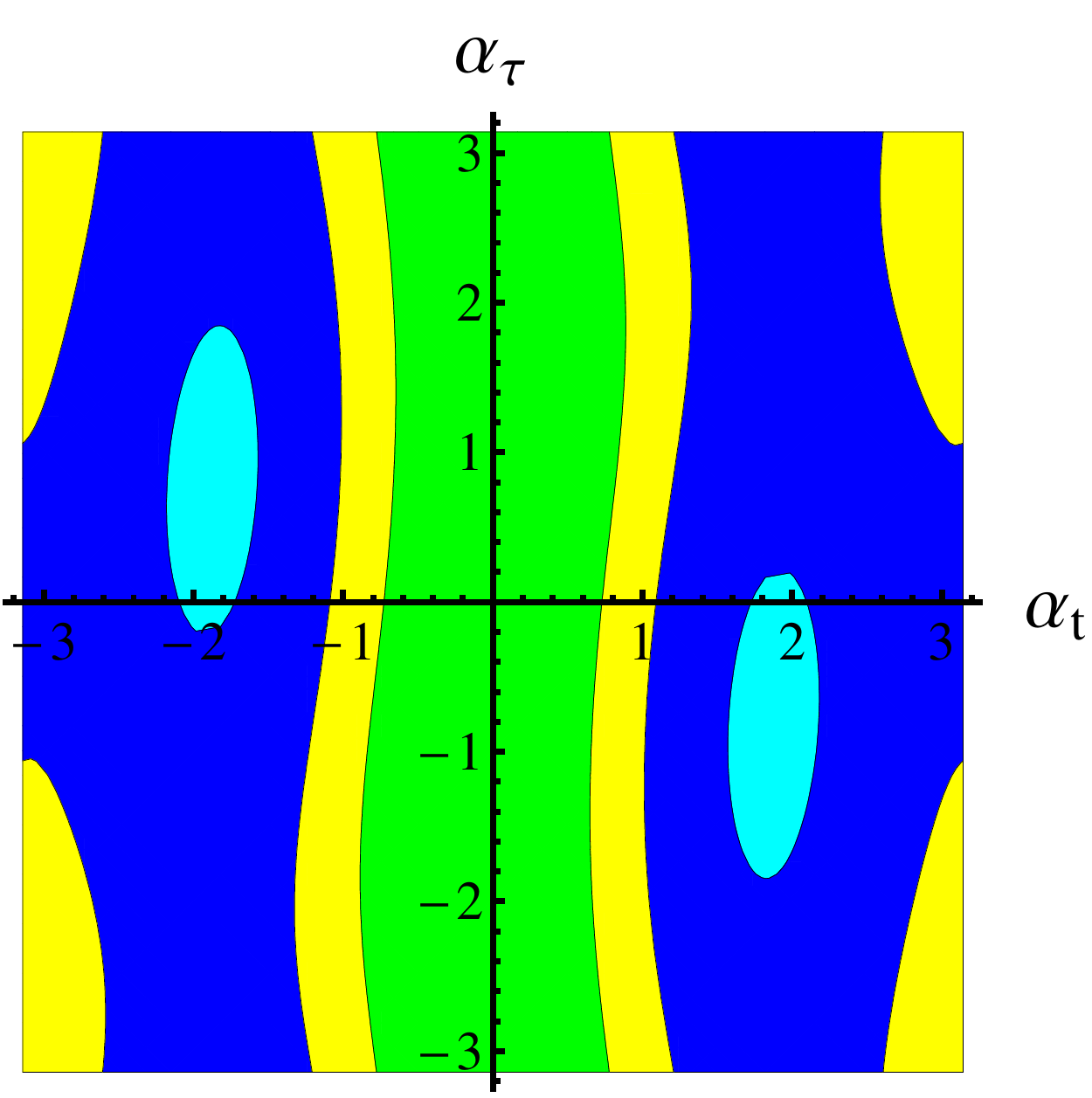}\quad\includegraphics[scale=0.4]{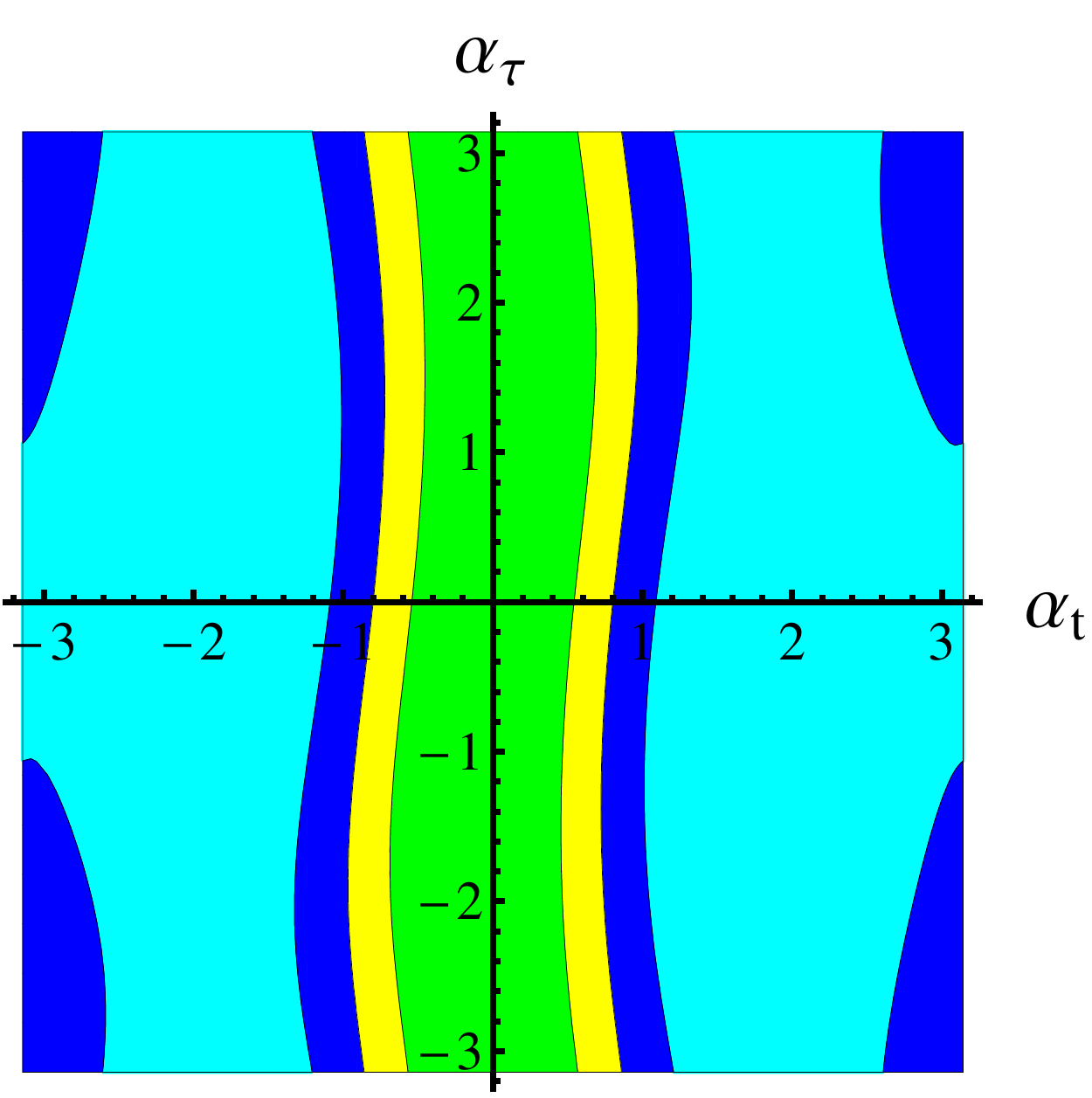}
\end{figure}

We show the $\textrm{Br}(\tau\rightarrow\mu\gamma)$ distributions in $\alpha_t-\alpha_{\tau}$ plane in
\autoref{futLee} with the boundaries set according to the sensitivity of Super $\tau$-charm factory, and
all the colored regions are for $\textrm{Br}(\tau\rightarrow\mu\gamma)<2.4\times10^{-9}$, thus case I in
\autoref{futurecase} would be disfavored.

In this scenario, the results for $\textrm{Br}(\tau\rightarrow\mu\gamma)$ cannot reach the sensitivity of SuperB
factory but they will reach the sensitivity of Super $\tau$-charm factory. If Super $\tau$-charm factory gave
negative results as case III in \autoref{futurecase}, it would give strict constraints on the Higgs couplings
as well that $|\alpha_t|\lesssim1$ would be favored but the constraints on $\alpha_{\tau}$ would be weak. While
if Super $\tau$-charm factory gave positive results as case II in \autoref{futurecase}, larger $|c_t|$ and $\alpha_t$
would be favored.
\subsection{LHC with Negative Result}
In this subsection we choose $(\sigma_h/\sigma_{h,\textrm{SM}})\textrm{Br}(h\rightarrow\mu\tau)=7.7\times10^{-4}$
as the LHC expected $95\%$ C.L. upper limit together with the replacement (\ref{rep}). In scenario I in \autoref{CBRE}
where the coupling strengths are close to those in SM, the predicted $\textrm{Br}(\tau\rightarrow\mu\gamma)
\lesssim(1-2)\times10^{-9}$; while in scenario II in \autoref{CBRE}, as the Lee model scenario, the predicted
$\textrm{Br}(\tau\rightarrow\mu\gamma)\lesssim(2-4)\times10^{-10}$.

\begin{figure}
\caption{$\textrm{Br}(\tau\rightarrow\mu\gamma)$ distributions in $\alpha_t-\alpha_{\tau}$ plane for for
$c_V=\Gamma_h/\Gamma_{h,\textrm{SM}}=1$, taking $|c_t|=1,1.2,1.5$ from left to right.
The green regions are for $\textrm{Br}(\tau\rightarrow\mu\gamma)<2\times10^{-10}$, the yellow regions are for $2\times10^{-10}\leq\textrm{Br}(\tau\rightarrow\mu\gamma)<5\times10^{-10}$, the blue regions are for
$5\times10^{-10}\leq\textrm{Br}(\tau\rightarrow\mu\gamma)<10^{-9}$, and the cyan regions are
for $10^{-9}\leq\textrm{Br}(\tau\rightarrow\mu\gamma)<2.4\times10^{-9}$.}\label{negsm}
\includegraphics[scale=0.4]{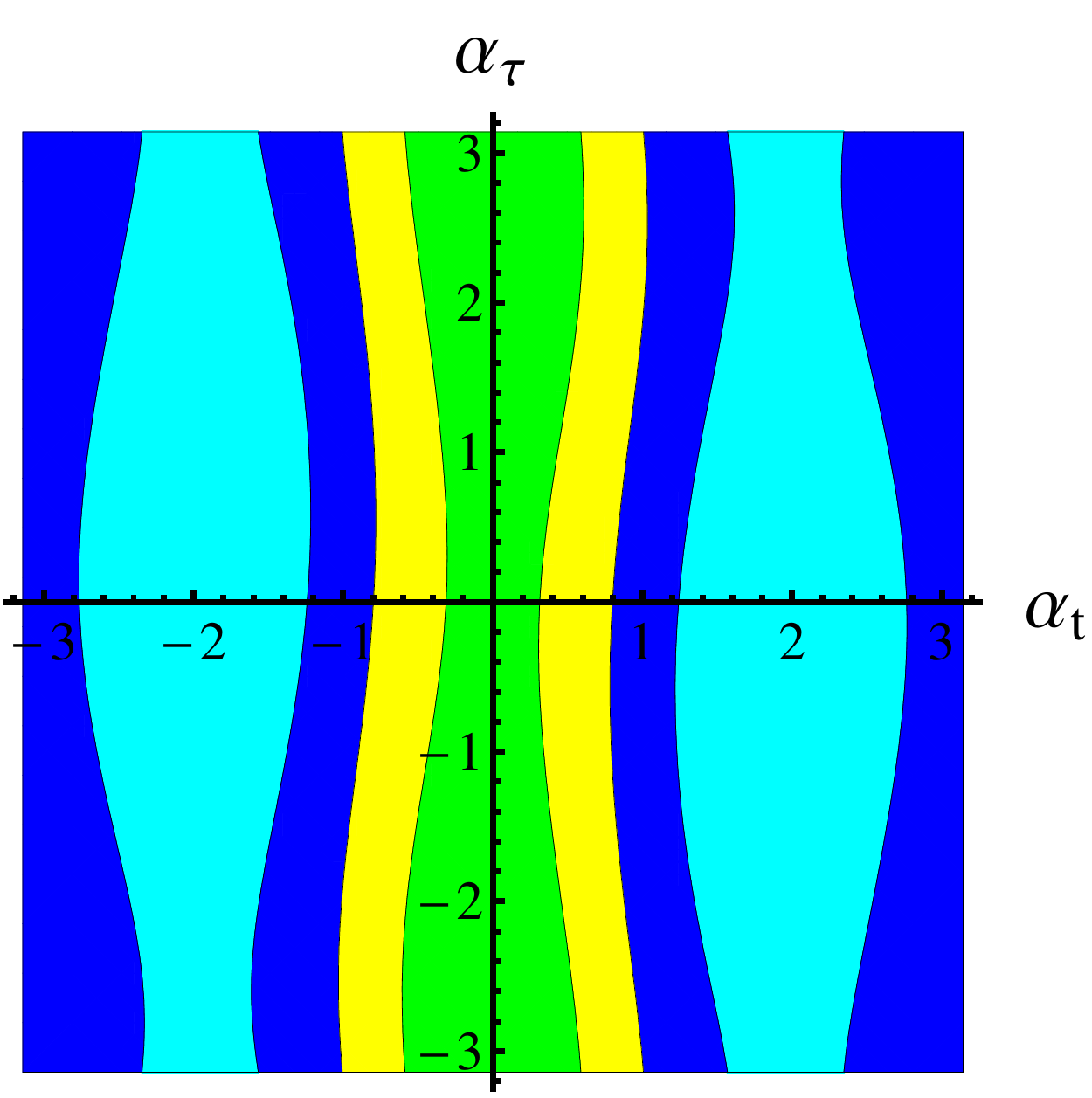}\quad\includegraphics[scale=0.4]{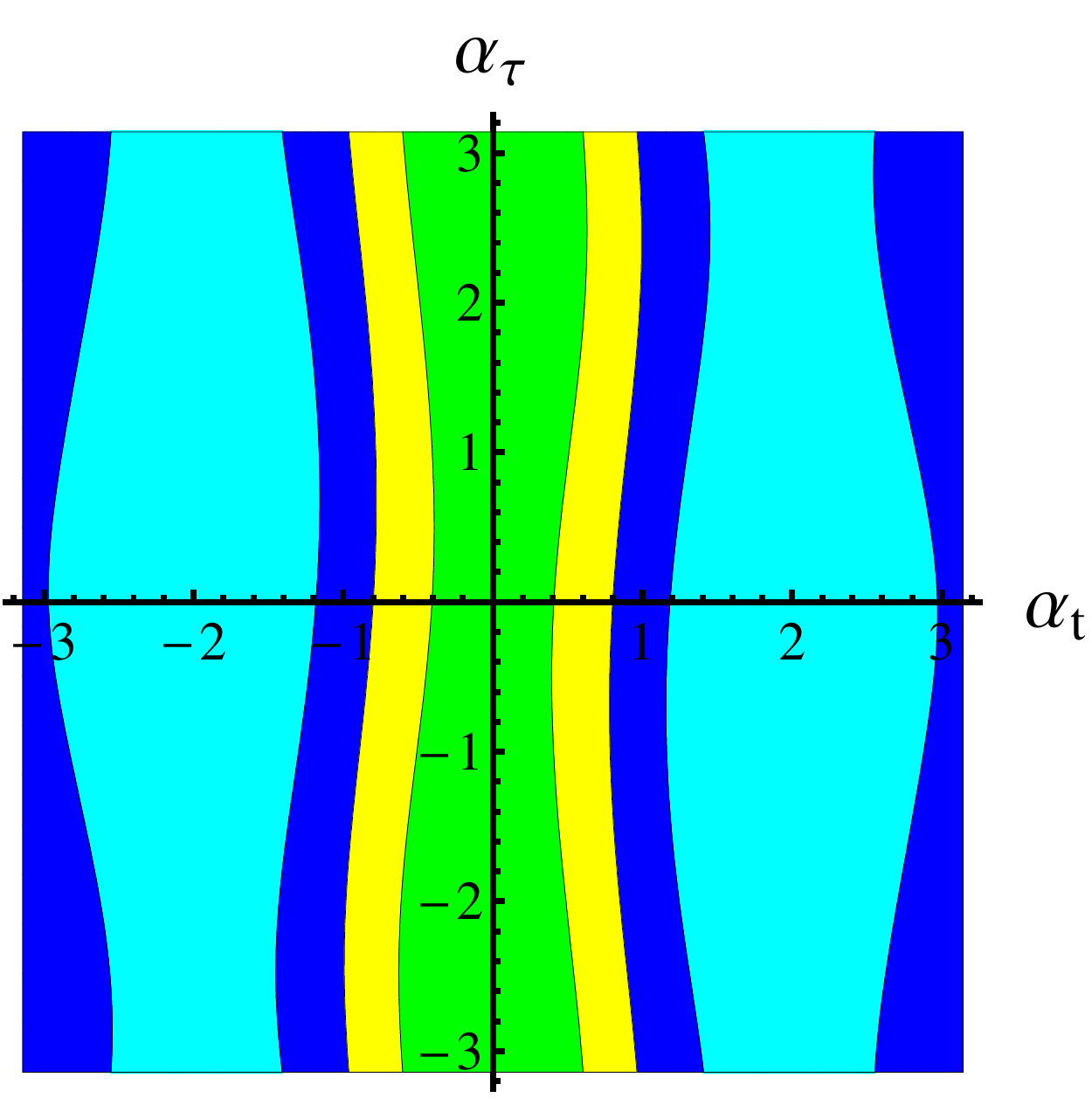}\quad\includegraphics[scale=0.4]{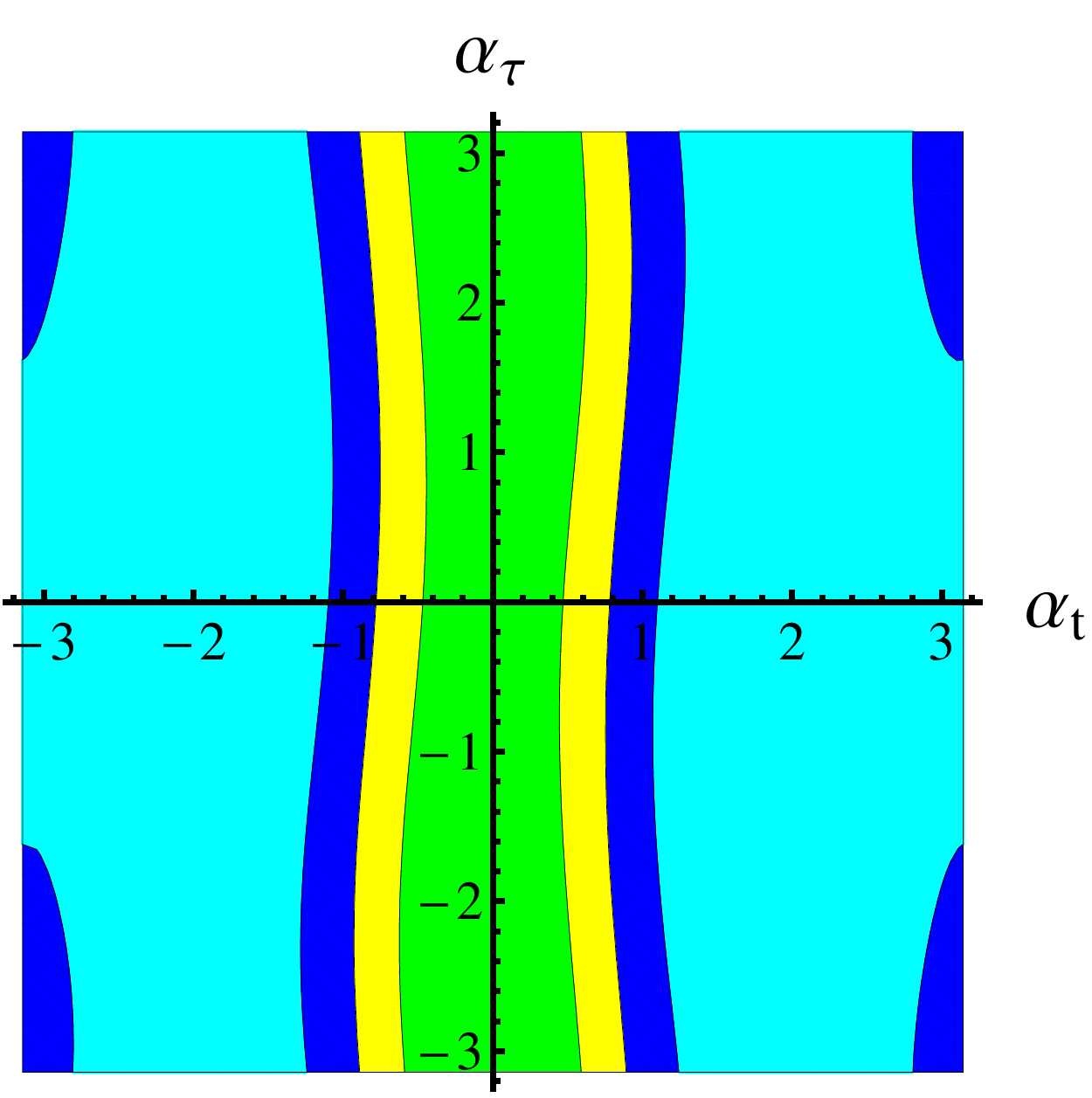}
\end{figure}
\begin{figure}
\caption{$\textrm{Br}(\tau\rightarrow\mu\gamma)$ distributions in $\alpha_t-\alpha_{\tau}$ plane for for
$c_V=0.5$ and $\Gamma_h/\Gamma_{h,\textrm{SM}}=0.3$, taking $|c_t|=1,1.2,1.5$ from left to right.
The green regions are for $\textrm{Br}(\tau\rightarrow\mu\gamma)<2\times10^{-10}$, the yellow regions are for $2\times10^{-10}\leq\textrm{Br}(\tau\rightarrow\mu\gamma)<5\times10^{-10}$.}\label{negLee}
\includegraphics[scale=0.4]{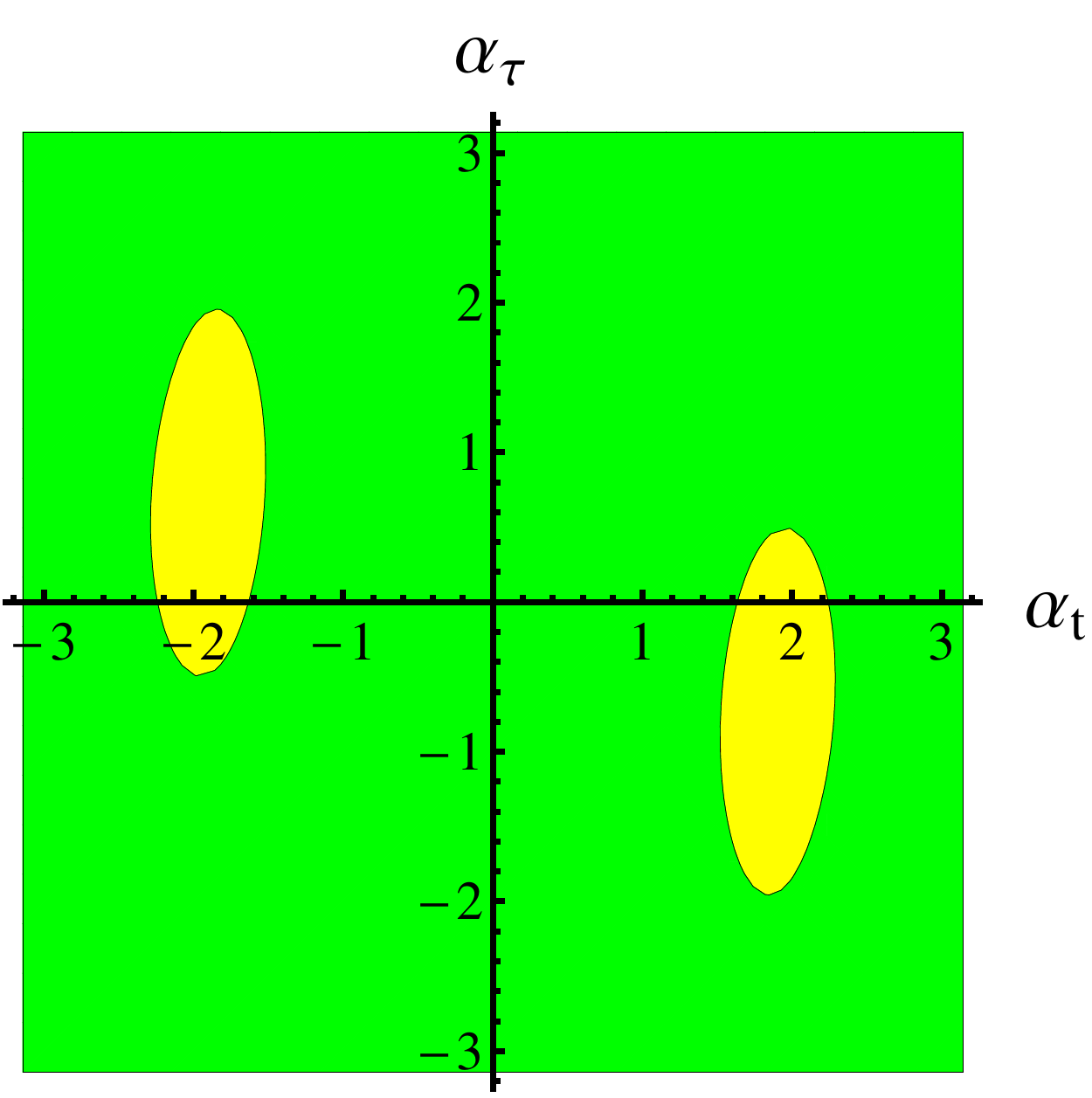}\quad\includegraphics[scale=0.4]{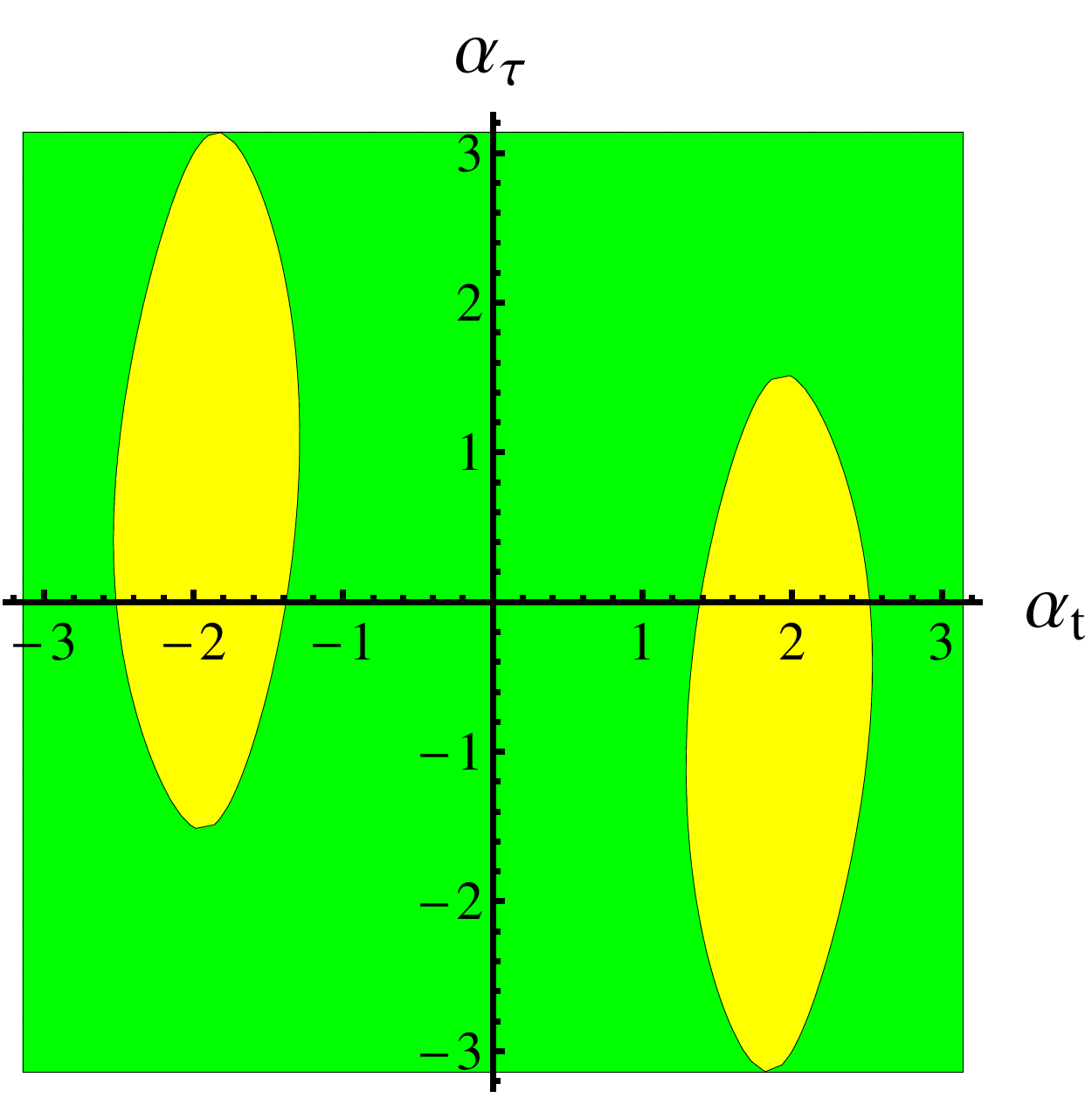}\quad\includegraphics[scale=0.4]{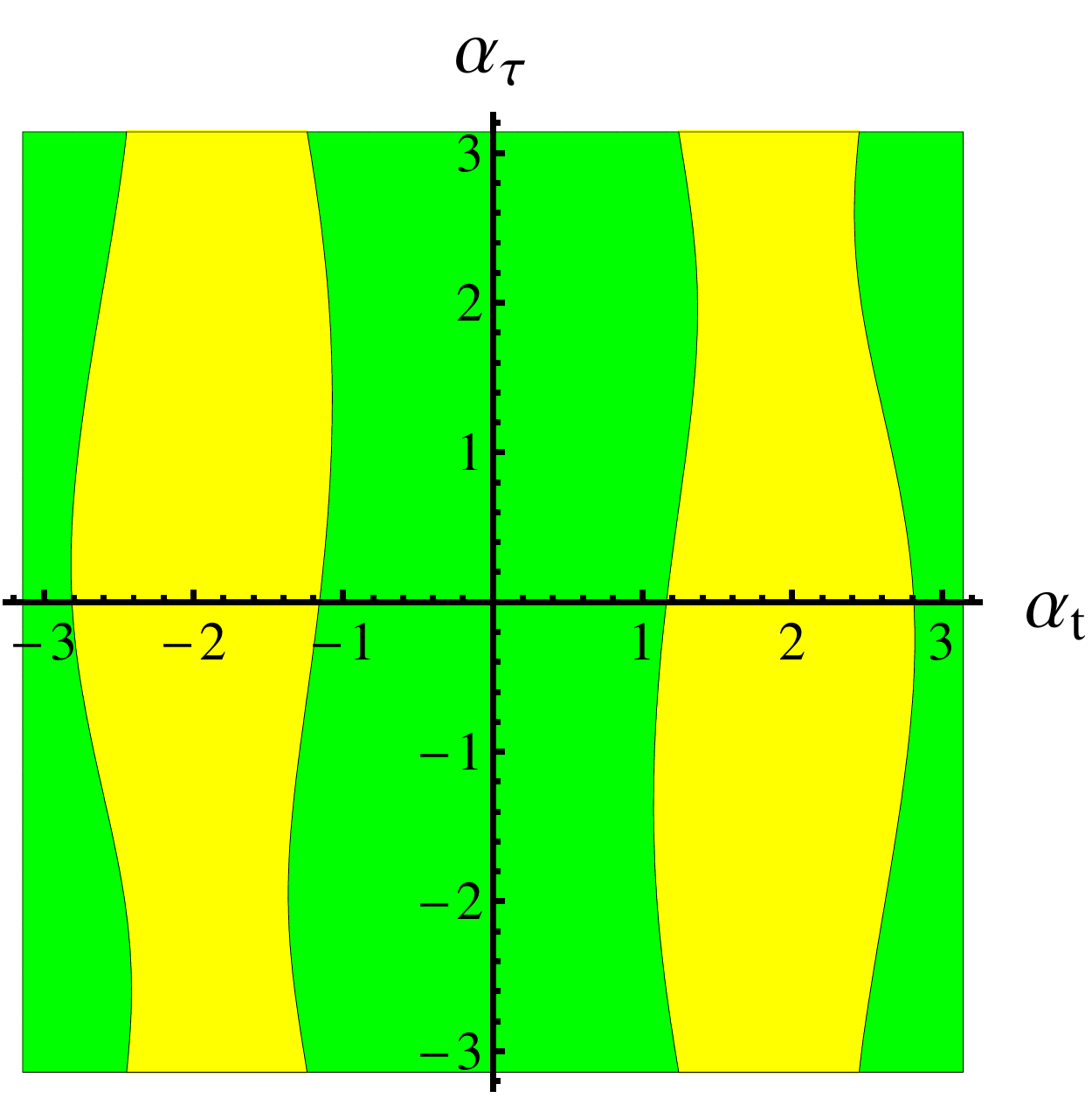}
\end{figure}
We should discuss the two scenarios separately. We show the $\textrm{Br}(\tau\rightarrow\mu\gamma)$ distributions
in $\alpha_t-\alpha_{\tau}$ plane in \autoref{negsm} for scenario I and in \autoref{negLee} for scenario II respectively.
If LHC gave negative results, the case I in \autoref{futurecase} cannot appear thus we focus on case II and III.
For scenario I, if Super $\tau$-charm factory gave negative results, $|\alpha_t|\lesssim(0.3-1)$ would be favored,
else the other regions would be favored. For scenario II, most regions are allowed for case III in \autoref{futurecase}
that both $e^+e^-$ colliders gave negative results.

\section{Conclusions and Discussions}
\label{CAD}
In this paper we discussed the Higgs-$\mu$-$\tau$ coupling induced LFV decay processes $h\rightarrow\mu\tau$ and
$\tau\rightarrow\mu\gamma$. For the later process, the branching ratio is also closely related to the
$ht\bar{t}$, $h\tau^+\tau^-$ and $hW^+W^-$ couplings.
We categorized the BSM into two scenarios, namely scenario I (II) with the Higgs coupling strengths close to
(far away from) those in the SM, and for the latter scenario we took the Lee model as an example.
We showed the possible numerical values of $\textrm{Br}(\tau\rightarrow\mu\gamma)$ for different cases
from \autoref{recent1} to \autoref{negLee}.

If the future LHC run gives positive results on $h\rightarrow\mu\tau$, different measurements on $\textrm{Br}(\tau\rightarrow\mu\gamma)$ at super B factory and super $\tau$-charm factory
 would distinguish
the two scenarios or imply the favored parameter choices. For case I in \autoref{futurecase}, with positive results
from both SuperB and Super $\tau$-charm factories, scenario I would be favored while scenario II would be disfavored
or even excluded. For typical parameter choices, see the blue or cyan regions in \autoref{futsm1} in details. For case
II in \autoref{futurecase}, with negative result from SuperB factory but positive result from  Super $\tau$-charm factory,
both scenarios would be allowed and some constraints would be given on the Higgs couplings. See blue and cyan regions in
\autoref{futsm2} and \autoref{futLee} for scenario I and II separately to find detail information on parameter choices.
For scenario I, $\alpha_{\tau}$ would be free for most cases, but regions near $(\alpha_t,\alpha_{\tau})=(0,\pm\pi)$ would
be disfavored for larger $|c_t|$ and $\textrm{Br}(h\rightarrow\mu\tau)$. For scenario II, $|\alpha_t|\gtrsim1$ would be
favored thus it implies large CP-violation in $ht\bar{t}$ coupling. For case III in \autoref{futurecase}, with negative
results from both SuperB and Super $\tau$-charm factories, scenario II would be more favored, but scenario I would not be
excluded. See green regions in \autoref{futsm2} and \autoref{futLee} for scenario I and II separately.

If the future LHC run gives negative results on $h\rightarrow\mu\tau$, case I in \autoref{futurecase} cannot be explained. If case I really
happened, we would need other models. For case II in \autoref{futurecase}, scenario I with $|\alpha_t|\gtrsim(0.5-1)$
would be favored, which implies large CP-violation in $ht\bar{t}$ coupling. While there would be almost no constraints on
$\alpha_{\tau}$. See \autoref{negsm} for details. For case III in \autoref{futurecase}, nothing about LFV are to be seen at
future colliders. Scenario I with $|c_t|\gtrsim(0.5-1)$ would be excluded, while other regions for both scenarios are allowed.

In \autoref{summary} we summarize the implications corresponded to all the six future possibilities depending on the
measurements of $\textrm{Br}(h\rightarrow \mu \tau)$ at the LHC and $\textrm{Br}(\tau\rightarrow\mu\gamma)$
at the super B factory and super tau charm factories.
\begin{table}[h]
\caption{Short summary on all the future possibilities depending on the measurements $\textrm{Br}(h\rightarrow \mu \tau)$ at the LHC and $\textrm{Br}(\tau\rightarrow\mu\gamma)$
at the super B factory and super $\tau$-charm factories and their corresponding implications.}\label{summary}
\begin{tabular}{|c|c|c|c|c|}
\hline
& \begin{tabular}{c} $\textrm{Br}(h\rightarrow\mu\tau)$ \\ @LHC\end{tabular}& \begin{tabular}{c} $\textrm{Br}(\tau\rightarrow\mu\gamma)$ \\ @SuperB\end{tabular} & \begin{tabular}{c} $\textrm{Br}(\tau\rightarrow\mu\gamma)$ \\ @Super $\tau$-charm\end{tabular} & Implications \\
\hline
P-I & Positive & Positive & Positive & \begin{tabular}{c}Scenario I favored;\\Scenario II excluded.\end{tabular}\\
\hline
P-II & Positive & Negative & Positive & \begin{tabular}{c}Both scenarios allowed;\\Scenario I with small $|\alpha_t|$ favored;\\Scenario II with large $|\alpha_t|$ favored.\end{tabular}\\
\hline
P-III & Positive & Negative & Negative & \begin{tabular}{c}Scenario II more favored;\\Scenario I with small $|\alpha_t|$ allowed.\end{tabular}\\
\hline
P-IV & Negative & Positive & Positive & Cannot be explained here.\\
\hline
P-V & Negative & Negative & Positive & \begin{tabular}{c}Scenario I with large $|\alpha_t|$ favored; \\Scenario II disfavored.\end{tabular}\\
\hline
P-VI & Negative & Negative & Negative & \begin{tabular}{c}Scenario I with large $|\alpha_t|$ disfavored;\\Other parameter regions allowed.\end{tabular}\\
\hline
\end{tabular}
\end{table}
With the help of future measurements on LFV processes $h\rightarrow\mu\tau$ and $\tau\rightarrow\mu\gamma$ at both high and low energy colliders,
for most cases, we would be able to distinguish different BSM scenarios or set constraints on Higgs couplings. P-IV in \autoref{summary} would
be strange. If it is really the case in the future, the Higgs induced LFV would not be the underlying reason. It would require other mechanism
beyond Higgs sector to generate large enough LFV processes such as $\tau\rightarrow\mu\gamma$.

\section*{Acknowledgement}
We thank Gang Li, Yuji Omura and Chen Zhang for helpful discussions. This work was supported in part by the Natural Science Foundation of
China (Grants No. 11135003 and No. 11375014).

\clearpage\end{CJK*}


\begin{thebibliography}{99}

\bibitem{CKM}N. Cabbibo, Phys. Rev. Lett. 10, 531 (1963); M. Kobayashi and T. Maskawa, Prog. Theor. Phys. 49, 652 (1973).
\bibitem{GIM}S. L. Glashow, J. Iliopoulos, and L. Maiani, Phys. Rev. D 2, 1285 (1970).
\bibitem{fcsm}W. J. Marciano and A. I. Sanda, Phys. Lett. B 67, 303 (1977).
\bibitem{PMNS}B. Pontecorvo, Zh. Eksp. Teor. Fiz. 33, 549 (1957) [Sov. Phys. JEPT 6, 429 (1957)];
 Z. Maki, M. Nakagawa, and S. Sakata, Prog. Theor. Phys. 28, 870 (1962).
\bibitem{no}K. A. Olive et al. (Particle Data Group), Chin. Phys. C 38, 090001 (2014).
\bibitem{exp1}K. Hayasaka et. al. (Belle Collaboration), Phys. Lett. B 666, 16 (2008);
 B. Aubert et. al. (BaBar Collaboration), Phys. Rev. Lett. 104, 021802 (2010).
\bibitem{exp2}J. Adam et. al. (MEG Collaboration), Phys. Rev. Lett. 110, 201801 (2013).
\bibitem{fut1}J. Brodzicka et al. (Belle Collaboration), Prog. Theor. Exp. Phys. 04D001 (2012), arXiv: 1212.5342.
\bibitem{fut2}T. Aushev et. al., Report No. KEK Report 2009-12, arXiv: 1002.5012.
\bibitem{fut3}SuperB Collaboration, Report No. INFN/AE-10/2, LAL-110, SLAC-R-952, arXiv: 1008.1541.
\bibitem{fut4}A. M. Baldini et. al. (MEG Collaboration), arXiv: 1301.7225.
\bibitem{Zhu2014hda}S.-H. Zhu, arXiv: 1410.2042.
\bibitem{dis1}ATLAS Collaboration, Phys. Lett. B 716, 1 (2012).
\bibitem{dis2}CMS Collaboration, Phys. Lett. B 716, 30 (2012).
\bibitem{pro}M. Flechl (for the ATLAS and CMS Collaborations), arXiv: 1503.00632.
\bibitem{CMS}CMS Collaboration, Report No. CMS-HIG-14-005 and CERN-PH-EP-2015-027, arXiv: 1502.07400.
\bibitem{ATLAS}ATLAS Collaboration, Report No. CERN-PH-EP-2015-184, arXiv: 1508.03372.
\bibitem{2hdm}G. C. Branco, P. M. Ferreira, L. Lavoura, M. N. Rebelo,
 M. Sher, and J. P. Silva, Phys. Rep. 516, 1 (2012).
\bibitem{2hdm3}J. D. Bjorken and S. Weinberg, Phys. Rev. Lett. 38, 622 (1977).
\bibitem{2hdm5}D. Aristizabal Sierra and A. Vicente, Phys. Rev. D 90, 115004 (2014).
\bibitem{2hdmanother}D. Das and A. Kundu, Phys. Rev. D 92, 015009 (2015).
\bibitem{2hdmanother2}F. J. Botella, G. C. Branco, M. Nebot, and M. N. Rebelo, Report No. IFIC-15-62, arXiv: 1508.05101.

\bibitem{2hdms}J. Heeck, M. Holthausen, W. Rodejohann and Y. Shimizu, Nucl. Phys. B 896, 281 (2015).
\bibitem{2hdms2}A. Crivellin, G. D'Ambrosio and J. Heeck, Phys. Rev. Lett. 114, 151801 (2015).
\bibitem{2hdms3}A. Crivellin, G. D'Ambrosio and J. Heeck, Phys. Rev. D 91, 075006 (2015).
\bibitem{our}Y.-N. Mao and S.-H. Zhu, Phys. Rev. D 90, 115024 (2014), arXiv: 1409.6844.
\bibitem{Lee}T. D. Lee, Phys. Rev. D 8, 1226 (1973).
\bibitem{other1}L. de Lima, C. S. Machado, R. D. Matheus, L. A. F. do Prado, JHEP 1511, 074 (2015).
\bibitem{other2}I. Dor$\check{\textrm{s}}$ner, S. Fajfer, A. Greljo, J. F. Kamenik, N. Ko$\check{\textrm{s}}$nik,
 and I. Ni$\check{\textrm{s}}$and$\check{\textrm{z}}$ic, JHEP 1506, 108 (2015).
\bibitem{other3}K. Cheung, W.-Y. Keung, P.-Y. Tseng, arXiv: 1508.01897.

\bibitem{phe1}B. Bhattacherjee, S. Chakraborty, and S. Mukherjee, arXiv: 1505.02688.
\bibitem{phe2}Y. Omura, E. Senaha, and K. Tobe, JHEP 05, 028 (2015), arXiv: 1502.07824.
\bibitem{tau1}G. Blankenburg, J. Ellis, and G. Isidori, Phys. Lett. B 712, 386 (2012).
\bibitem{tau2}R. Harnik, J. Kopp, and J. Zupan, JHEP 03, 026 (2013).

\bibitem{phe3}C.-J. Lee and J. Tandean, JHEP 04, 174 (2015).
\bibitem{CS}T. P. Cheng and M. Sher, Phys. Rev. D 35, 3484 (1987).
\bibitem{STC}E. Levichev, Phys. Part. Nucl. Lett. 5, 554 (2008).
\bibitem{STC2}A. V. Bobrov and A. E. Bondar, Nucl. Phys. B (Proc. Suppl.) 253-255, 199 (2014).
\bibitem{zfactory} J. P. Ma and C. H. Chang, Sci. China (Phys., Mech. and Astro.) 53, 1947-1948 (2010).
\bibitem{width}The LHC Higgs Cross Section Working Group, Report No. CERN-2013-004, arXiv: 1307.1347.
\bibitem{BZ}S. M. Barr and A. Zee, Phys. Rev. Lett. 65, 21 (1990); 65, 2920 (1990).
\bibitem{loop}S. Davidson and G. Grenier, Phys. Rev. D 81, 095016 (2010).
\bibitem{loop2}D. Chang, W.-S. Hou, and W.-Y. Keung, Phys. Rev. D 48, 217 (1993).
\bibitem{Kandn}J. Kopp and M. Nardecchia, JHEP 1410, 156 (2014), arXiv: 1406.5303.
\end{thebibliography}
\end{document}